%% file: main.tex
\documentclass[a4paper, 11pt]{article}
\usepackage{jheppub}

\pdfoutput=1

\usepackage[utf8]{inputenc}
\usepackage{epsf, array, xcolor}
\usepackage{graphicx}
\usepackage{bbold}
\usepackage{amsmath, amssymb, mathtools}
\usepackage{epigraph}
\usepackage{textcomp}
\usepackage{color}
\usepackage{slashed}
\usepackage{graphics,psfrag}
\usepackage{graphicx,psfrag}
\usepackage{cancel}
\usepackage{float}
\usepackage{url}
\usepackage{pbox}
\usepackage{feynmp-auto}
\usepackage{etoolbox}
\usepackage{fontawesome}
\AtBeginEnvironment{tabular}{\scriptsize}

\newcommand{\mhsm}{m_{h}}
\newcommand{\hsm}{h}

\title{\boldmath Higgs Decays at NLO in the SMEFT}

\author[a]{Luigi Bellafronte,}
\author[b]{Sally Dawson,}
\author[b]{Clara Del Pio,}
\author[c]{Matthew Forslund,}
\author[d]{ and Pier Paolo Giardino}

\emailAdd{lbellafronte@fsu.edu}
\emailAdd{dawson@bnl.gov}
\emailAdd{cdelpio@bnl.gov}
\emailAdd{mforslund@princeton.edu}
\emailAdd{pier.giardino@uam.es}

\affiliation[a]{Physics Department, Florida State University, Tallahassee, FL 32306-4350, USA}
\affiliation[b]{High Energy Theory Group, Physics Department, 
    Brookhaven National Laboratory, Upton, NY 11973, USA}
\affiliation[c]{Princeton Center for Theoretical Science, Princeton University, Princeton, NJ, 08544 USA}
\affiliation[d]{Departamento de Física Teórica and Instituto de Física Teórica UAM/CSIC, 
    Universidad\\ Autónoma de Madrid, Cantoblanco, 28049, Madrid, Spain}

\abstract{The calculation of precise predictions for Higgs decays is a necessary ingredient for determining Higgs properties at the LHC and future colliders.  We compute all two- and three- body Higgs decays at next-to-leading order (NLO)  in both QCD and electroweak interactions using the dimension-6 Standard Model Effective Field Theory (SMEFT). Results for four-body Higgs decays that are accurate to NLO QCD/electroweak order in the SMEFT are obtained using the narrow width approximation. Our results are contained in a flexible Monte Carlo program,  {\sc NEWiSH}, that is publicly available and we illustrate the impact of the NLO electroweak corrections for HL-LHC, Tera-Z, and Higgstrahlung projections. 

{\sc NEWiSH} is publicly available through \href{https://gitlab.com/mforslund/newish}{GitLab~\faGitlab}.}

\makeatletter
\gdef\@fpheader{}
\makeatother

\begin{document}
\preprint{\quad} IFT-UAM/CSIC-26-4
\maketitle

\section{Introduction}\label{sec:intro}
\input{intro.tex}

\section{SMEFT NLO Calculation}\label{sec:smeft_nlo}
\input{smeft}
\section{Processes}\label{sec:processes}
\input{process.tex}
\section{Results}\label{sec:results}
\input{results}

\section{Conclusions}\label{sec:conclusions}
\input{conc.tex}
\section*{Acknowledgments}
We would like to thank Marvin Schnubel for naming inspiration of the {\sc NEWiSH} code and Andrea Visibile for numerical comparisons.
S. D. and C.D.P. are supported
by the U.S. Department of Energy under Contract No. DE-
SC0012704. P.P.G. is supported by the Ramón y Cajal grant~RYC2022-038517-I funded by MCIN/AEI/10.13039/501100011033 and by FSE+, and by the Spanish Research Agency (Agencia Estatal de Investigación) through the grant IFT Centro de Excelencia Severo Ochoa~No~CEX2020-001007-S. The work of L.B. is supported in part by the U.S.
Department of Energy under Grant No. DE-SC0010102 and by the College of Arts and Sciences of Florida State University. L.B. thanks the Technische Universität München (TUM) for the hospitality and the Excellence Cluster ORIGINS which is funded
by the Deutsche Forschungsgemeinschaft (DFG, German Research Foundation)
under Germany´s Excellence Strategy – EXC-2094 – 390783311 for
partial support during the completion of this work. Digital data including partial widths and the code {\sc NEWiSH} is provided at a \href{https://gitlab.com/mforslund/newish}{GitLab~\faGitlab}~repository.
\bibliography{refs.bib}
\bibliographystyle{utphys}
\newpage
\appendix
\section{Code}\label{sec:Code}
\input{code.tex}
\section{Numerical results for \texorpdfstring{$\hsm \to f\bar{f}$ and $\hsm \to VV^\prime$}{h to ff and h to VV}}\label{sec:app_num}
For completeness, here we include numerical results for $\hsm \to f\bar{f}$ and $\hsm \to VV'$ using the numerical inputs in Sec.~\ref{sec:inputs}.
All results here are also available as data files at our NEWiSH GitLab repository~\cite{GITLAB:newish}.
Additionally, for $\hsm \to q\bar{q}$ we include results for $\overline{\mathrm{MS}}$ quark masses as well.
\input{hff.tex}

\input{hvv.tex}
\section{Projected Sensitivities at Tera-Z}
\input{terasens.tex}


\end{document}

%% file: intro.tex
A precise understanding of Higgs boson interactions is central to the physics programs of present and future colliders.  
Currently, measurements of single Higgs production are in good agreement with theoretical predictions both for production and decay processes~\cite{CMS:2022dwd,ATLAS:2022vkf}.
With future high luminosity LHC (HL-LHC) measurements, however, the experimental uncertainties will be significantly reduced~\cite{ATL-PHYS-PUB-2025-018}, requiring increasingly precise theoretical predictions.\footnote{See~\cite{Huss:2025nlt,Proceedings:2019vxr,Spira:2016ztx} for relevant summaries of current state of the art calculations in the Standard Model.}  
Deviations from Standard Model (SM) predictions in the Higgs sector are further restricted by the lack of evidence for direct production of new particles which typically correspond to extensions of the Standard Model.  
It is thus reasonable to assume that any new physics corresponds to new interactions and/or new particles that are relevant at scales much larger than the electroweak scale.

The search for new physics can be systematically pursued using the Standard Model Effective Field Theory (SMEFT)~\cite{Brivio:2017vri}.  
This approach assumes that all Beyond the Standard Model (BSM) physics is at a very high scale  compared to the weak scale and that the weak scale fields are those of the Standard Model  interacting through an $SU(3)\times SU(2) \times U(1)$ gauge symmetry. 
Furthermore, the SMEFT approach  assumes that the electroweak symmetry breaking is linearly realized as in the SM and that the Higgs boson is part of an $SU(2)$  doublet.
The new physics can then be described as an expansion around the SM Lagrangian,
\begin{equation}\label{eq:SMEFTexp}
\mathcal{L}=\mathcal{L}_\textrm{SM}+\sum_{i,d}{\frac{C_i^d}{\Lambda^{d-4}}}O_i^d\, ,
\end{equation}
where $\Lambda$ is the scale of the new physics and $O_i^d$ are operators of dimension-$d$ containing only SM fields.
All information about the nature of the new physics is contained in the coefficient functions, $C_i^d$. 

The theory we have just described is a consistent field theory that admits a well-defined renormalization procedure, valid order by order in powers of $\Lambda$. 
In other words, once a specific power of $\Lambda$ has been chosen, corrections to SM processes can be computed in a straightforward manner in terms of $C_i^d/\Lambda^{d-4}$ and to any desired order in the loop expansion by consistently truncating higher powers of $\Lambda$ that may appear in loop diagrams or when constructing physical observables. 
In our study we truncate the expansion at dimension-6 and we neglect the dimension-5 operators which generate lepton number violating interactions. 
At dimension-6, observables depend only on the ratio $C/\Lambda^2$ and there is no independent sensitivity to the scale $\Lambda$\footnote{We drop the superscript $"6"$ in the remainder of this paper.}.

QCD corrections to one-loop order in the dimension-6 SMEFT are automated~\cite{Brivio:2020onw,Degrande:2020evl} and can be calculated for all experimentally relevant processes.  
Higher order electroweak (EW) corrections in the SMEFT, however, need to be performed on a case-by-case basis.  
Numerous examples of electroweak dimension-6 SMEFT one-loop calculations exist:  $W$ and $Z$ pole observables~\cite{Bellafronte:2023amz,Dawson:2019clf,Biekotter:2025nln}, Drell-Yan production~\cite{Dawson:2018dxp,Dawson:2021ofa,ElFaham:2024egs}, $e^+e^-\rightarrow Z \hsm$~\cite{Asteriadis:2024xts,Asteriadis:2024xuk}, and many 2- and 3- body Higgs decays~\cite{Bellafronte:2025jbk,Cullen:2020zof,Cullen:2019nnr,Gauld:2016kuu,Dawson:2018pyl,Dedes:2019bew,Dawson:2018liq,Dedes:2018seb,Hartmann:2015aia,Dawson:2024pft,Martin:2023fad,Corbett:2021cil}.  
In this work, we summarize NLO results for all two- and three- body Higgs decays at one-loop in the CP conserving dimension-6 SMEFT, presenting new NLO results for the decays $\hsm\rightarrow W f{\overline{f}}'$,  $\hsm\rightarrow Z q {\overline{q}}$ and   
$\hsm \rightarrow Z \nu\bar{\nu}$.
Importantly, our results are contained in a public code that produces both total Higgs decay widths and distributions at one-loop in the SMEFT including all QCD and electroweak corrections.    
This code is part of a significant effort by many researchers to compute the  Higgs production and decay channels that are measured at the LHC to one-loop electroweak order in the SMEFT to facilitate a global fit that is accurate to this order.\footnote{An important and still missing piece for a complete NLO description of SMEFT processes are the two-loop  SMEFT renormalization group equations (RGE).  The bosonic two-loop contributions are known~\cite{Born:2024mgz}.} 
The calculations are also crucial for the study of Higgstrahlung and $Z$ pole observables at future $e^+e^-$ colliders.  

At one-loop order in the SMEFT, a plethora of operators typically contribute to observables, with the exact number depending on the underlying flavor assumptions.  
For example, $Z$ and $W$  pole observables (EWPOs) at leading order (tree) depend on 10 dimension-6 operators, while at next-to-leading order (one-loop), the number increases to 32 in the scenario where flavor is ignored. 
Assuming a $U(2)^5$ scenario, for example, the number of operators contributing to EWPOs  increases to 93~\cite{Greljo:2022cah,Bellafronte:2023amz}. 
In this work, we combine previous one-loop electroweak SMEFT results for $e^+e^-\rightarrow Z \hsm$~\cite{Asteriadis:2024xts,Asteriadis:2024xuk} with our  NLO results for $\hsm\rightarrow X$, where $X$ is a two- or three- body final state. 
All possible flavor interactions are included in the calculation, however we drop contributions proportional to the off-diagonal elements of the Cabibbo-Kobayashi-Maskawa (CKM) matrix. 
To recover the four fermion final state we use combinations of full results and the narrow width approximation (NWA), as described in Sec. \ref{sec:h24f}.

In Sec.~\ref{sec:smeft_nlo}, we present a detailed discussion of the SMEFT one-loop electroweak calculation of Higgs decays and the dipole subtraction used to regulate the real photon and gluon emission contributions. 
We then discuss each Higgs decay individually in Sec.~\ref{sec:processes}, including numerical results for the widths.
In Sec.~\ref{sec:results}, we present some numerical examples of the impact of our work including partial NLO projections for the sensitivities for Higgs processes at the HL-LHC. 
We give a comparison of sensitivities at Tera-Z and HL-LHC to selected operators that occur at loop level and hence would be omitted in tree level SMEFT studies.  
As a case study of the impact of NLO electroweak corrections in the SMEFT, we demonstrate the sensitivity of the Higgs singlet model to NLO corrections and emphasize the impact of the model assumptions on the projections.  
The use of our public code {\sc NEWiSH}~\cite{GITLAB:newish} for the study of one-loop dimension-6 SMEFT operators in Higgs decays is detailed in an appendix.

%% file: smeft.tex
We begin by establishing our conventions and discussing the details of our calculations, first broadly and then individually for each process. 
The starting point is  the dimension-6 SMEFT Lagrangian~ Eq. \eqref{eq:SMEFTexp} in the Warsaw basis~\cite{Grzadkowski:2010es,Dedes:2017zog}. 
We do not assume any specific flavor structure for the operators that can appear in our calculation. 
However, the CKM matrix is taken to be diagonal, to restrict the flavour structures that appear in any given process.

We obtain the amplitudes that contribute to the virtual one-loop corrections using the chain {\sc FeynRules}~\cite{Alloul:2013bka}$\rightarrow ${\sc FeynArts}~\cite{Hahn:2000kx}$\rightarrow $ {\sc FeynCalc}~\cite{Shtabovenko:2023idz}. 
The resulting Feynman integrals, which we calculate using $D=4-2\epsilon$ dimensional regularization, are expressed in terms of Passarino-Veltman functions~\cite{Passarino:1978jh}, whose analytical and numerical expressions we obtain from  {\sc Package-X}~\cite{Patel:2016fam} and {\sc Collier}~\cite{Denner:2016kdg}.
For all processes, we take as input parameters
\begin{equation}
    M_W, \ M_Z, \ G_\mu,\  \mhsm, \ m_t,\ \alpha_S\, ,
\end{equation}
where the masses are on-shell, and the Fermi constant $G_\mu$ is determined from the decay of the muon. 
The masses of leptons and quarks (excluding the top quark), are generally taken to be zero. 
However, for the Higgs decays into light fermions, $\hsm\to f\bar{f}$, we also include the masses of second and third generation quarks when their impact is sufficiently large. 
We leave the details to the respective sections. 

Phase space integration is handled via a public Fortran code {\sc NEWiSH}~\href{https://gitlab.com/mforslund/newish}{\faGitlab}~\cite{GITLAB:newish} that we describe in App.~\ref{sec:Code}.
The calculation of the three-body decays mirrors that of~\cite{Asteriadis:2024xts,Asteriadis:2024xuk,Dawson:2024pft} for $\ell^+\ell^-Z$, extended to include the $q{\overline{q}}Z$, $\nu {\overline{\nu}}Z$, and $f\bar{f}^\prime W^\pm$ processes.
For three-body decay processes $\hsm\rightarrow f\bar{f}^\prime V$ ($V=W,Z$), we take all fermions massless except for the top-quark.\footnote{With the exception of in some differential distributions -- see Sec.~\ref{sec:llZ} for details.}

For $\hsm\rightarrow f\bar{f}$ decays, in order to have a consistent SM contribution, light fermion masses need to be included.
In this context, adopting an $\overline{\textrm{MS}}$ rather than an on-shell definition of the input masses makes a significant numerical difference, as we discuss in Sec.~\ref{sec:msbar}.
For the $\hsm\to f \bar{f}$ decays, since the mediating vector bosons can be on-shell, we adopt the complex mass scheme for the treatment of the unstable particles~\cite{Denner:2006ic,Dittmaier:2009cr}.

To obtain predictions for the four-body decays $\hsm\rightarrow (f_1 \bar{f}_2)( f_3 \bar{f}_4)$ at NLO in the SMEFT, we use the narrow width approximation (NWA) with our $\hsm\rightarrow f\bar{f} V$ NLO decay calculation and the known results for $W$ and $Z$ decays at NLO~\cite{Dawson:2019clf, Bellafronte:2023amz}.
At lowest order (LO), we calculate the full four-body process with no approximation and employ the complex mass scheme, which ensures a gauge-invariant description of the off-shell vector bosons. 
We discuss the narrow width approximation in more detail in Sec.~\ref{sec:h24f}.\footnote{Some of the pitfalls of the NWA at tree level in the SMEFT are discussed in~\cite{Brivio:2019myy}.}

\subsection{Renormalization}

We  calculate our results employing two different renormalization conditions for the counterterms. 
In the first approach, we use the on-shell (OS) scheme for the SM parameters, while the ${\overline{\rm MS}}$ scheme is used for the renormalization of SMEFT operators.
The second approach is very similar to the first, however the masses of the light quarks (i.e all excluding the top quark), are calculated in the ${\overline{\rm MS}}$ scheme. 
As usual, we define the OS quantities in terms of the masses of the bosons and fermions, with the addition of the Fermi constant $G_\mu$. 
We note that $e$ is not an independent parameter of the model, ($e$ is defined as the coupling of the electron to the photon in SMEFT).
It is convenient to express the electromagnetic coupling in terms of our input parameters as
\begin{align}
\label{eq:charge}
\begin{split}
     e^2 &= {G_\mu}\sqrt {2}
     \left(1+ X_H  \right) 4 M_W^2 \left( 1- \frac{M_W^2}{M_Z^2}\right)\\
     &-\frac{2M_W^3} {M_Z^2\Lambda^2}\biggl\{M_W C_{\phi D} +{4}\sqrt{M_Z^2-M_W^2}C_{\phi W B}\biggr\} +{\cal{O}}\biggl(\frac{1}{\Lambda^4}\biggr) \, ,
\end{split}
\end{align}
where at dimension-6
\begin{align}
    X_H\equiv \frac{1}{\sqrt{2}G_\mu\Lambda^2}\biggl\{
    C_{ll}[1221]- C_{\phi l}^{(3)}[11]-C_{\phi l}^{(3)}[22]\biggr\} \, .
    \label{eq:xdef}
\end{align}
The indices in the square brackets are generation indices.

We fix the tadpole renormalization by setting tadpole contributions to zero \cite{Denner:2019vbn}. This approach corresponds to equating the renormalized vacuum expectation value (VEV) of the Higgs field to the minimum of the renormalized scalar potential. The relation between the renormalized VEV $v$ and $G_\mu$ is then given by, 
\begin{align}
\label{eq:gmu:smeft}
\begin{split}
    \sqrt{2}G_\mu (1+X_H)&=\frac{1}{v^2}(1+\Delta r)\, .
\end{split}
\end{align}

The function  $\Delta r$~\cite{Sirlin:1980nh, Marciano:1980pb} has been calculated in the SMEFT at dimension-6 in \cite{Dawson:2018pyl}. Notice that, due to our choice of tadpole scheme, $\Delta r$ is a gauge-dependent quantity. We use the gauge-dependence of our intermediate calculations to check the correctness of the final result. Specifically, we explicitly check the gauge independence of the result for the SM calculation and for specific parts of the SMEFT calculation, as checking the gauge invariance of the full SMEFT calculation is  impractical due to the large size of the amplitudes involved.

The renormalization conditions for the masses of the vector bosons $V$ ($V=W,Z$) are set by the definitions
\begin{align}
    M_V^2=M_{0,V}^2-\delta M_V^2\, ,
\end{align}
which relate the bare mass $M_{0,V}^2$ to the physical mass $M_V^2$. $\delta M_V^2$ are the mass counterterms and, at one-loop, are defined by the relations
\begin{align}
  \delta M_V^2   =\text{Re}(\Pi_{T,VV}(M_V^2))\, ,
\end{align}
where $\Pi_{T, VV}(M_V^2)$ are the transverse parts of the one-loop corrections to the two-point functions evaluated at the physical mass $M_V^2$. Notice that, following the standard definition, we take the real part of the two-point functions. We will see in Sec. \ref{sec:hffsection} that we need to use a complex mass scheme for the $\hsm\to f{\overline{f}}$ decays due to a divergence in the real corrections to those processes  when the $Z$ boson propagator goes on-shell. However, since vector bosons do not contribute at LO in those decays we do not need to alter our renormalization procedure.

For the Higgs boson, the renormalization condition is very similar, and we define
\begin{align}
    \mhsm^2=m_{0,h}^2-\delta \mhsm^2\, , \, \delta \mhsm^2   =\text{Re}(\Pi_{\hsm\hsm}(\mhsm^2)) \, 
\end{align}
where, again, $m_{0,\hsm}$ and $\mhsm$ are the bare and physical masses of the Higgs boson, respectively, and $\Pi_{\hsm\hsm}(\mhsm^2)$ is the one-loop correction to the Higgs boson two-point function evaluated at the physical Higgs mass. 

For the fermions it is convenient to first define the fermionic two-point function, which we write as
\begin{align}
\Sigma(p)=\frac{1-\gamma_5}{2}(\slashed{p} \Sigma_{p,L}+\Sigma_{m,L})+\frac{1+\gamma_5}{2}(\slashed{p} \Sigma_{p,R}+\Sigma_{m,R}),
\end{align}
where $p$ is the momenta of the fermion. The relation between the bare fermionic mass $m_{0,f}$ and physical mass $m_{f}$ then becomes
\begin{align}
m_{f}^2=m_{0,f}^2-\delta m_f^2\, ,
\end{align}
where
\begin{align}
\delta m_f^2= m_f^2 (\Sigma_{p,L}+\Sigma_{p,R})+m_f  (\Sigma_{m,L}+\Sigma_{m,R})\, .
\end{align}

These formulas are valid in both the OS and $\overline{\rm MS}$ schemes; the only distinction is whether one subtracts the full function or only its divergent part, respectively.

Regarding the wave-function renormalization, we have the following definitions:
\begin{align}
&    \delta  Z_{\gamma\gamma} =  - \left.  \frac{d \Pi_{\gamma\gamma}}{d p^2} \right|_{p^2 = 0} \, , 
    \delta  Z_{VV} = - \left.  \frac{d \Pi_{VV}}{d p^2} \right|_{p^2 = M_V^2} \, ,  
    \delta  Z_\hsm = - \left. \frac{d \Pi_{\hsm\hsm}}{d p^2} \right|_{p^2 = \mhsm^2} \, ,\nonumber \\
& \delta Z_{f,L}= -\left. \Sigma_{p,L} + 2 \frac{d}{d p^2}(m_f^2 \Sigma_{m,L}+ m_f \Sigma_{p,L} )\right|_{p^2 = m_f^2}\, , 
\nonumber \\ & \delta Z_{f,R}= -\left. \Sigma_{p,R} + 2 \frac{d}{d p^2}(m_f^2 \Sigma_{m,R}+ m_f \Sigma_{p,R} )\right|_{p^2 = m_f^2}\, .
\end{align}

Additional wave-function renormalization counterterms come from the requirement that off-diagonal $\gamma-Z$ mixing terms vanish at the poles, from which we have
\begin{align}
    \delta Z_{\gamma Z} = -2 \frac{\Pi_{\gamma Z}(M_Z^2)}{M_Z^2} \, , \quad  
    \delta Z_{Z\gamma} = 2  \frac{\Pi_{\gamma Z}(0)}{M_Z^2} \, ,
\end{align}
where $\Pi_{\gamma Z}$ is the contribution from the two-point Feynman diagram which mixes photons and $Z$ bosons. Notice that $\delta Z_{\gamma\gamma}$ receives contributions from light quarks that do not vanish in the massless limit. These are related to the hadronic contribution to the vacuum polarization, defined as
$\Delta\alpha^{(5)}_{\mathrm{had}}\equiv e^2\left(\Pi^{(5)}_{\gamma\gamma}(M_Z^2)-\Pi^{(5)}_{\gamma\gamma}(0)\right)$,
where $\Pi^{(5)}_{\gamma\gamma}$ denotes the light-quark contribution to the photon two-point function calculated at two different scales, $\Delta\alpha^{(5)}_{\mathrm{had}}$ is extracted from experimental data, and $e^2$ in our scheme is defined in Eq.~\ref{eq:charge}.
It is interesting to note that the corresponding leptonic contributions are logarithmically divergent, and we regulate these divergences using the lepton masses.

Finally, as advertised at the beginning of the section, we renormalize the bare dimension-6 SMEFT coefficients, $C_{0,i}$, in the ${\overline{\rm MS}}$ scheme
\begin{align}
    C_i(\mu) = C_{0,i}- \frac{1}{2\hat\epsilon}  \frac{1}{16\pi^2}\gamma_{ij}C_j(\mu) \, ,
\end{align}
where $\hat\epsilon^{-1}\equiv\epsilon^{-1}-\gamma_E+\log (4\pi)$, $\mu$ is the renormalization scale, and $\gamma_{ij}$
are the elements of the anomalous dimension matrix~\cite{Jenkins:2013zja,Jenkins:2013wua,Alonso:2013hga,Alonso:2014zka}, that is
\begin{align}
    16\pi^2 \, \frac{\textrm{d}C_i (\mu)}{\textrm{d}\ln\mu} = \gamma_{ij}C_j(\mu) \, .
\end{align}

Coefficients in our numerical expressions for the widths are evaluated at the scale $\mu=\mhsm$ and we do not include the RGE running of the coefficients needed to match to models at the scale $\Lambda$~\cite{terHoeve:2025gey,Bartocci:2024fmm}.

\subsection{Real emission}\label{sec:real_emission}

Infrared (IR) divergences in the virtual amplitudes are canceled by real photon or real gluon emission, $\hsm\rightarrow X + \gamma/g$.
We treat these using standard dipole subtraction techniques, as described in~\cite{Catani:1996vz,Dittmaier:1999mb,Catani:2002hc,Denner:2019vbn}.
Depending on the process, we require a combination of massive~\cite{Dittmaier:1999mb,Catani:2002hc} and massless~\cite{Catani:1996vz,Denner:2019vbn} dipole subtraction.
Explicitly, for the real emission contribution, we have
\begin{equation}\label{eq:subtraction}
    \Gamma_\text{Real} = \frac{1}{2\mhsm} \int \mathrm{dPS}_n  \left( |\mathcal{A}_\mathrm{R}|^2 - |\mathcal{A}_\mathrm{sub}|^2\right) + \int \mathrm{d}\Gamma_\mathrm{sub}
\end{equation}
where $|\mathcal{A}_\mathrm{R}|^2$ is the amplitude squared for the real emission process $H\rightarrow X + \gamma/g$,  expanded to $\mathcal{O}(1/\Lambda^2)$.
Since all of our splittings are of the form $X \rightarrow \gamma/g(p_i) + X(p_j)$ with spectator particle momentum $p_k$ and final-state emission only, the subtraction kernel $|\mathcal{A}_\mathrm{sub}|^2$ is given by (see Sec.~5 of~\cite{Catani:2002hc})
\begin{align}\label{eq:subkern}
    |\mathcal{A}_\mathrm{sub}|^2 =&-8\pi \alpha_S C_F \sum_{\langle jk\rangle } |\mathcal{A}_\mathrm{LO}(p_j\rightarrow \tilde{p}_{ij},p_k\rightarrow \tilde{p}_k) |^2\frac{ 1}{(p_i+p_j)^2-m_j^2} \\ & \times\left\{\frac{2}{1-\tilde{z}_j(1-y_{ij,k})}-\frac{\tilde{v}_{ij,k}}{v_{ij,k}} \left[ 1+\tilde{z}_j + \frac{m_{j}^2}{p_i \cdot p_j}\right] \right\}\, ,
\end{align}
where the sum is over emitter-spectator pairs $\langle j k\rangle$, $m_{i\ldots k}^2 = (p_i + \ldots + p_k)^2$, and $\mathcal{A}_\mathrm{LO}$ is the leading order amplitude for $\hsm\to X$ expanded to $\frac{1}{\Lambda^2}$.
For real photon emission one replaces $4\pi \alpha_S C_F \rightarrow e^2 Q_j Q_k$ with $Q_j$ the electric charge of particle $j$ and $e$ the QED gauge coupling expanded to $\frac{1}{\Lambda^2}$ in terms of the input parameters, as defined in Eq. \eqref{eq:charge}.
The new variables in Eq.~\eqref{eq:subkern} are given by
\begin{align}
    \tilde{z}_j &= \frac{p_i\cdot p_k}{(p_i\cdot p_k+p_j\cdot p_k)} \, , \qquad y_{ij,k} = \frac{p_{i}\cdot p_j}{p_i \cdot p_j + p_i \cdot p_k + p_j \cdot p_k} 
\end{align}
\begin{align}
    \tilde{v}_{ij,k} &= \frac{\lambda^{1/2}(m_{ijk}^2,m_j^2,m_k^2)}{m_{ijk}^2-m_j^2-m_k^2} \, ,
\end{align}
\begin{align}
v_{ij,k} &= \frac{\sqrt{[2m_k^2+(m_{ijk}^2-m_j^2-m_k^2)(1-y_{ij,k})]^2-4m_{ijk}^2 m_k^2}}{(m_{ijk}^2-m_j^2-m_k^2)(1-y_{ij,k})} \, ,
\end{align}
where $\lambda(a,b,c)$ is the usual Källén triangle function.
The LO amplitude $|{\mathcal{A}}_\mathrm{LO}(\tilde{p}_{ij},\tilde{p}_k)|^2$ in Eq. \eqref{eq:subkern} is evaluated with momenta $\tilde{p}_{ij}$ and $\tilde{p}_k$ given by
\begin{align}
\begin{split}
    \tilde{p}_{k}^\mu &= \frac{\lambda^{1/2} (m_{ijk}^2,m_{j}^2,m_k^2)}{\lambda^{1/2} (m_{ijk}^2,m_{ij}^2,m_k^2)} \left(p_k^\mu - \frac{p_{ijk} \cdot p_k}{m_{ijk}^2} p_{ijk}^\mu\right) + \frac{p_{ijk}^2 - m_k^2-m_j^2}{2m_{ijk}^2} p_{ijk}^\mu\\
    \tilde{p}^\mu_{ij} &= p_{ijk}^\mu-\tilde{p}_k^\mu
\end{split}
\end{align}
where $p_{ijk}^\mu \equiv p_i^\mu + p_j^\mu + p_k^\mu$.
In the massless limit, $m_j,m_k\rightarrow0$, these formulas reproduce the massless case in~\cite{Denner:2019vbn}.  We will however need the full mass dependence for the $\hsm\rightarrow f\bar{f}$ and $\hsm\rightarrow f\bar{f}' W^\pm$ processes described below.

The subtracted piece is  integrated analytically using dimensional regularization and added back as $\int \mathrm{d}\Gamma_\text{sub}$, given explicitly by
\begin{equation}
    \mathrm{d\Gamma}_\mathrm{sub} = \frac{1}{2\mhsm}\mathrm{dPS}_{n-1}  \frac{\alpha_S}{2\pi}C_F \sum_{\langle jk \rangle}|\mathcal{A}_\mathrm{LO}|^2 \left(2 I^\mathrm{eik}_+ + I^\mathrm{coll}_{gQ,k}\right) \, ,
\end{equation}
where the integrals $ I^\mathrm{eik}_+$ and $I^\mathrm{coll}_{gQ,k}$ are given in Eqs. 5.34 and 5.35 in~\cite{Catani:2002hc}.

It should be kept in mind that the dipole subtraction formulae written here are valid only for inclusive observables.
For a non-inclusive observable such as $d\Gamma /dm_{\ell\ell}$ for $\hsm\rightarrow \ell^+\ell^- Z$, the lepton mass $m_\ell$ enters as a regulator and additional terms appear. 
See Sec.~\ref{sec:llZ} for details.

The total decay width is then the sum of the real and virtual pieces described above.

\subsection{Inputs}\label{sec:inputs}

In the numerical results that follow, we use a renormalization scale $\mu = \mhsm$ to define the observables and a SMEFT scale $\Lambda = 1$~TeV.  
We take as our inputs~\cite{Freitas:2023iyx}
\begin{align}
\begin{split}
    M_W  &= 80.352~\textrm{GeV} \\
    M_Z  &= 91.1535~\textrm{GeV} \\
    G_\mu &= 1.16638\times 10^{-5}~\textrm{GeV}^{-2} \\
    \mhsm &= 125.1~\textrm{GeV}\\
    m_t &= 172.76~\textrm{GeV}\\
    \alpha_S(\mhsm) &= 0.1188 \\
    \Delta \alpha^{(5)}_\textrm{had} &= 0.02768 \, .
\end{split}
\end{align}
The vector boson masses 
denote here the pole definitions~\cite{Bardin:1988xt,Freitas:2023iyx}
\begin{equation}
\label{eq:MZMZMZ}
    M_V=\frac{M_V^{\exp}}{\sqrt{1+\left({\Gamma_V^{\exp}}/{M_V^{\exp}}\right)^2}} \, ,
\end{equation}
and correspond to experimental values of $M_W^{\text{exp}}=80.379~\textrm{GeV}$, $M_Z^{\text{exp}}=91.1876~\textrm{GeV}$, $\Gamma_W^{\text{exp}} = 2.085$ GeV, $\Gamma_Z^{\text{exp}} = 2.4952$ GeV. 
For the $\hsm\rightarrow f\bar{f}$ processes, we take the additional on-shell inputs~\cite{LHCHiggsCrossSectionWorkingGroup:2016ypw}
\begin{align}
\begin{split}
    m_\tau &= 1.777~\textrm{GeV}\\
    m_\mu &= 0.1057~\textrm{GeV}\\
    m_e &= 0.511\times 10^{-3}~\textrm{GeV}\\
    m_b &= 4.92~\textrm{GeV}\\
    m_c &= 1.51~\textrm{GeV}\\
    m_s &= 0.1~\textrm{GeV}\, ,
\end{split}
\end{align}
where only masses $m_f \geq m_{f_i}$ are taken to be nonzero for each $\hsm\rightarrow f_i \bar{f}_i$ channel.
See the description of each individual process below for more detail.
For $\hsm \to q\bar{q}$, we additionally show numerical results using $\overline{\mathrm{MS}}$ input values~\cite{ParticleDataGroup:2024cfk},
\begin{align}
\begin{split}
    m_b^{\overline{\mathrm{MS}}} (m_b) &= 4.183~\textrm{GeV}\\
    m_c^{\overline{\mathrm{MS}}} (m_c) &= 1.273~\textrm{GeV}\\
    m_s^{\overline{\mathrm{MS}}} (2 \; {\rm GeV})&= 0.0935~\textrm{GeV}\, .
\end{split}
\end{align}

%% file: process.tex
In this section, we discuss the different two- and three- body processes that contribute to the Higgs width at NLO QCD and EW in the dimension-6 SMEFT and our implementation in {\sc NEWiSH}.\footnote{
We neglect a handful of ultrarare decays described in~\cite{dEnterria:2025rjj} that are also vanishingly small in the SMEFT at $\mathcal{O}(1/\Lambda^2)$.}
With the exceptions of $\hsm\to VV^\prime$ with $VV^\prime\in(gg,\gamma\gamma,Z\gamma)$ and $\hsm\to ggZ$ which are loop induced in the SM, the LO plus one-loop virtual
amplitudes for the decay processes $\hsm\to X$ are given by 
\begin{equation}
    \mathcal{A}(\hsm\to X) = \mathcal{A}^{(0)}_\text{SM} + \frac{1}{\Lambda^2}\mathcal{A}^{(0)}_\text{EFT} + \frac{1}{(4\pi)^2}\mathcal{A}^{(1)}_\text{SM} + \frac{1}{(4\pi \Lambda)^2}\mathcal{A}^{(1)}_\text{EFT} + \ldots
    \label{eq:expdef}
\end{equation}
where we truncate everywhere at $\mathcal{O}(\frac{1}{16\pi^2\Lambda^2})$. 
The LO plus virtual $\hsm \to X$ amplitudes- squared are then simply 
\begin{align}\label{eq:ampsq_generic}
\begin{split}
    |\mathcal{A}_{\text{Virt}}|^2 =& |\mathcal{A}^{(0)}_\text{SM}|^2 + \frac{2}{\Lambda^2} \mathrm{Re}\left(\mathcal{A}_\text{SM}^{(0)*}\mathcal{A}_\text{EFT}^{(0)}\right)\\&+ \frac{2}{(4\pi)^2} \mathrm{Re}\left(\mathcal{A}_\text{SM}^{(0)*}\mathcal{A}_\text{SM}^{(1)}\right)+ \frac{2}{(4\pi \Lambda)^2} \mathrm{Re}\left(\mathcal{A}_\text{SM}^{(0)*}\mathcal{A}_\text{EFT}^{(1)}+\mathcal{A}_\text{SM}^{(1)*}\mathcal{A}_\text{EFT}^{(0)}\right)+ \ldots
\end{split}
\end{align}
where the first line is LO and the second is part of the NLO correction \footnote{We should emphasize that we always truncate the amplitude squared $|\mathcal{A}|^2$ at $\mathcal{O}(1/\Lambda^2)$.
Consistently including all $\mathcal{O}(1/\Lambda^4)$ contributions to an observable requires both double insertions of dimension-6 operators and the inclusion of dimension-8 operators to ensure UV pole cancellation and gauge invariance~\cite{Deutschmann:2017qum,Asteriadis:2022ras}.}.

The total NLO widths also require contributions from real photon and gluon emission, as discussed in Section~\ref{sec:real_emission}.
The relevant amplitudes squared for  the $\hsm \to X + \gamma/g$ processes are given by
\begin{equation}
    |\mathcal{A}(\hsm\to X+\gamma/g)|^2 = |\mathcal{A}_{X+\gamma/g, \text{SM}}^{(0)}|^2+\frac{2}{\Lambda^2}\mathrm{Re}\left[\mathcal{A}_{\text{SM},X+\gamma/g}^{(0)*} \mathcal{A}_{ \text{EFT},X+\gamma/g}^{(0)} \right] +\ldots
\end{equation}

For each process, we  present numeric results for the total NLO partial width $\Gamma_\textrm{NLO}(\hsm\to X)$ with the inputs of Sec.~\ref{sec:inputs} in the form $\Gamma_\textrm{NLO} = \Gamma_\textrm{LO} + \delta\Gamma_\textrm{NLO}$, where $\Gamma_\textrm{LO}$ is the leading order contribution  in the dimension-6 SMEFT and $\delta\Gamma_\textrm{NLO}$ is the total NLO correction.
We drop coefficients with negligibly small contributions, though full numerical results for each process including all contributions can be found in the GitLab repository~\cite{GITLAB:newish}.
Note that the three-body decays $h\to f\bar{f}\gamma$~\cite{Han:2017yhy,Han:2018juw,Corbett_2022} and $h\to f\bar{f}g$~\cite{Gao:2016jcm,Fox:2025cuz} are included as components of the inclusive $h\to f\bar{f}$ decay rates, but are not considered as exclusive channels since the required cuts are frame dependent and depend on the production mode.
We leave their inclusion as exclusive modes to future work.
We now consider each process in more detail.

\subsection{\texorpdfstring{$\hsm\to Z f {\overline{f}}$}{H to Zff}}

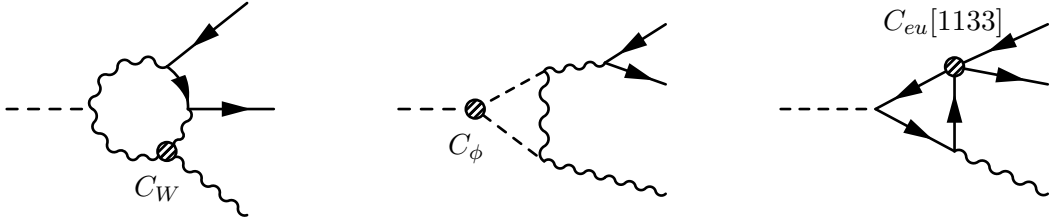
\begin{figure}
    \centering
    \begin{fmffile}{nlo_box}
\begin{fmfgraph*}(100,80)
    \fmfleft{i1,i2,i3}
    \fmfright{o1,o2,o3}
    \fmftop{t1,t2}
    \fmfbottom{b1,b2}
    \fmf{dashes}{i2,v1} 
    \fmf{phantom,t=0.9}{v1,v2}
    \fmf{fermion}{v2,o2}
    \fmffreeze
    \fmf{phantom,t=1}{vtl,i3}
    \fmf{phantom,t=0.6}{vtl,vtr} 
    \fmf{phantom,t=1}{vtr,o3} 
    \fmf{phantom,left=0.26,t=1.5}{v1,vtl,vtr} 
    \fmf{fermion,left=0.26,t=1.5}{vtr,v2} 
    \fmffreeze
    \fmf{boson,left=0.57,t=1,l.d=3}{v1,vtr} 
    \fmf{phantom,t=1}{vbl,i1} 
    \fmf{phantom,t=0.6}{vbl,vbr} 
    \fmf{phantom,t=1}{vbr,o1} 
    \fmf{phantom,right=0.26,t=1.5}{v1,vbl,vbr}   \fmf{boson,right=0.26,t=1.5,l.d=3}{vbr,v2}
    \fmffreeze
    \fmf{boson,right=0.57,t=1}{v1,vbr}
    \fmfblob{7}{vbr}
    \fmf{boson,t=0}{vbr,o1}
    \fmf{fermion,t=0}{o3,vtr}
    \fmfv{l.d=10,l.a=-110,l=\color{black}{$C_W$}}{vbr}
  \end{fmfgraph*}
\end{fmffile} \hspace{1cm}
\begin{fmffile}{nlo_cphi}
  \begin{fmfgraph*}(100,80)
    \fmfstraight
    \fmfright{o4,o3,o2,o1}
    \fmfleft{i2,h,i1}
    \fmftop{top}
    \fmfbottom{bot}
    \fmf{dashes,tension=1}{h,t1}
    \fmf{dashes,tension=.6}{t1,t3}
    \fmf{boson,tension=.6}{t3,t2}
    \fmf{dashes,tension=.6}{t2,t1}
    \fmf{phantom,tension=.8}{i1,t2}
    \fmf{phantom,tension=0.08}{i1,t1}
    \fmf{phantom,tension=.8}{t3,i2}
    \fmf{boson,tension=2.6}{t2,v1}
    \fmf{boson,tension=2.6}{v2,t3}
    \fmf{boson,tension=2.6}{v2,o4}
    \fmfshift{8 down}{o1}
    \fmfshift{ 4 down}{o2}
    \fmfshift{ 4 up}{o3}
    \fmfshift{8 up}{o4}
    \fmf{fermion,tension=1.3}{o1,v1,o2}
    \fmfblob{7}{t1}
    \fmfv{l.d=10,l.a=-110,l=\color{black}{$C_\phi$}}{t1}
  \end{fmfgraph*}\hspace{1cm}
\end{fmffile}
\begin{fmffile}{nlo_ceett}
  \begin{fmfgraph*}(100,80)
    \fmfstraight
    \fmfright{o4,o3,o2,o1}
    \fmfleft{i2,h,i1}
    \fmftop{top}
    \fmfbottom{bot}
    \fmf{dashes,tension=1}{h,t1}
    \fmf{fermion,tension=.6}{t1,t3,t2,t1}
    \fmf{phantom,tension=.8}{i1,t2}
    \fmf{phantom,tension=.8}{t3,i2}
    \fmfshift{8 down}{o1}
    \fmfshift{ 4 down}{o2}
    \fmfshift{ 4 up}{o3}
    \fmfshift{8 up}{o4}
    \fmf{fermion,tension=1}{o1,t2,o2}
    \fmf{phantom,tension=1}{o3,t3}
    \fmf{boson,tension=1}{o4,t3}
    \fmfblob{7}{t2}
    \fmfv{l.d=12,l.a=110,l=\color{black}{$C_{eu}[1133]$}}{t2}
  \end{fmfgraph*}
\end{fmffile}
\caption{Representative diagrams for  virtual contributions to the $\hsm\to f\bar{f}Z$ decay process in the SMEFT.}
  \label{fig:h_to_ffZ}
\end{figure}

In the SM, the $\hsm\to Zf\bar{f}$ processes are contained in the full $\hsm\to 4f$ decays, which are known at NLO QCD and EW order in the SM~\cite{Boselli:2015aha,Kaur:2023eyv,Bredenstein:2006rh,Bredenstein:2006ha}. 
The SM corrections most relevant for $\hsm\to Zf\bar{f}$ are conveniently implemented in {\sc Prophecy4f}~\cite{Denner:2019fcr}, {\sc Hto4l}~\cite{Boselli:2015aha}, and {\sc HDECAY}~\cite{Djouadi:2018xqq}.

In the SMEFT, only $\hsm\to Z\ell^+\ell^-$ ($l=\mu, e$) is known to one-loop, as described in detail in~\cite{Dawson:2024pft}. 
Here we extend that calculation to include $\hsm\to Z\tau^+\tau^-$, $\hsm\to Z\nu \bar{\nu}$ and $\hsm\to Z q\bar{q}$ using the same methodology. Sample diagrams contributing to the virtual contributions are shown in Fig. \ref{fig:h_to_ffZ}.
As discussed above, we consider all fermions massless except for $m_t$, although the lepton masses enter as a regulator in non-inclusive observables in $\hsm\to Z\ell^+\ell^-$.
We do not include the effects of finite $W$ and $Z$ widths in either the real emission or the virtual amplitudes since the $Z$ appears as an external state~\footnote{In a full $\hsm\to 4f$ calculation one could impose the complex mass scheme as we do in $\hsm\to f\bar{f}$. We do this for our leading order predictions in Sec.~\ref{sec:h24f}, but it is inconsistent at NLO when using the narrow width approximation.}.

\subsubsection{\texorpdfstring{$\hsm\to Z \ell^+\ell^-$}{H to Zll}}\label{sec:llZ}

The $\hsm\to Z \ell^+\ell^- $ process gives us access to semi-leptonic and fully leptonic $\hsm\to 4f$ decays in the narrow width approximation.
Since $\hsm\to 4\ell$ is particularly important at the LHC, we have also implemented $m_{\ell\ell}$ cuts and the non-inclusive differential distribution $d\Gamma/d{m}_{\ell\ell}$ for this channel\footnote{A similar approach may be used to implement non-inclusive observables for other $\hsm\to Vf\bar{f}$ decays as well. 
However, these are less phenomenologically relevant and more susceptible to neglected resummation effects from parton showering.
We  leave this for future work.}.

To include non-inclusive observables, an additional term is required in the dipole subtraction procedure.
This term is described in~\cite{Dawson:2024pft} and we write it here for completeness.
In the notation of~\cite{Denner:2019vbn}, this amounts to using in Eq.~\eqref{eq:subtraction}
\begin{equation}
    |\mathcal{A}_\mathrm{sub}|^2 = 2 e^2 \frac{p_{\ell^+} \cdot p_{\ell^-}}{(p_{\ell^+} \cdot p_\gamma)( p_{\ell^-} \cdot p_\gamma)}
\end{equation}
\begin{align}
\begin{split}
    \int \mathrm{d\Gamma}_\mathrm{sub}^{\hsm\to Z\ell^+\ell^-} =&  
 \bigg[\int \mathrm{d}\widetilde{\mathrm{PS}}_3 |\mathcal{A}_\text{LO}|^2 \int_0^1 dz \left\{G^{(\text{sub})}({\widetilde{m}}^2_{\ell\ell})\delta{(1-z)} 
 +\left[\bar{\mathcal{G}}_\mathrm{MR}
 ({\widetilde{m}}_{\ell\ell}^2,z)\right]_+\right\} \\ &
    \times\Theta_\text{cut}(p_{\ell^-} = z\widetilde{p}_{\ell^-},p_{\ell^+} = \widetilde{p}_{\ell^+},p_\gamma = (1-z)\widetilde{p}_{\ell^-})\bigg]+(p_{\ell^-} \leftrightarrow p_{\ell^+}) \, ,
\end{split}
\end{align}
where ${\widetilde{m}}_{\ell\ell}^2 = (\widetilde{p}_{\ell^-}+\widetilde{p}_{\ell^+})^2$ is $z$-independent and the tilde over $\mathrm{d}\widetilde{\mathrm{PS}}_3$ is to indicate that the integration is over the momenta $\widetilde{p}_{\ell^\pm}$.
The plus distribution is defined in the usual way, $\int_0^1 [f(z)]_+g(z) dz = \int_{0}^{1} f(z)[g(z)-g(1)]dz$, and the function $\Theta_\mathrm{cut}$ indicates which momenta are subject to phase space cuts. 
The functions $G^{(\mathrm{sub})}$ and $\bar{\mathcal{G}}_\mathrm{MR}$ are given by~\cite{Denner:2019vbn}
\begin{align}
\begin{split}
    G^{(\mathrm{sub})}({\widetilde{m}}_{\ell\ell}^2) =& \Gamma(1+\epsilon)\left(\frac{4\pi\mu^2}{{\widetilde{m}}_{\ell\ell}^2}\right)^\epsilon \left(\frac{1}{\epsilon^2}+\frac{3}{2\epsilon}\right)+\frac{7}{2}-\frac{\pi^2}{3}\\
    \bar{\mathcal{G}}_\text{MR}({\widetilde{m}}_{\ell\ell}^2,z) &= \hat{P}_{ff}\left[\ln{\left(\frac{{\widetilde{m}}_{\ell\ell}^2}{m_\ell^2}\right)}+\ln{z}-1\right]+(1+z)\ln{\left(1-z\right)}+1-z \, ,
\end{split}
\end{align}
where $\hat{P}_{ff} =(1+z^2)/(1-z)$. 
For inclusive observables, the contribution from $\bar{\mathcal{G}}_\mathrm{MR}$ vanishes and one recovers the massless limit of the subtraction described in Section~\ref{sec:real_emission}.
IR singularities from the virtual diagrams cancel with those in $G^{(\mathrm{sub})}$, while the physical lepton mass $m_\ell$ in $\bar{\mathcal{G}}_\mathrm{MR}$ acts as a regulator for the additional collinear singularities in non-inclusive observables.

Denoting by $i$ the flavour index of the lepton $\ell_i \in (e,\mu,\tau)$, our result for the inclusive $\hsm\to Z\ell_i^+\ell_i^-$ partial width is given by (note that the index $i$ is not summed over),
\begin{align}
\begin{split}
    \Gamma_\textrm{LO}&(\hsm\to Z\ell_i^+\ell_i^-) = 2.95 \times 10^{-6} \ \textrm{GeV} \\
    &+\left(\frac{1 \ \textrm{TeV}}{\Lambda}\right)^2\bigg(-0.255 C_{\phi B}+0.358 C_{\phi \square}+0.0596 C_{\phi D}+0.0802 C_{\phi W}\\
    &-0.153 C_{\phi WB}+0.358 C_{ll}[1221]-0.0207 C_{\phi e}[ii]+0.0258 C_{\phi l}^{(1)}[ii]-0.358 C_{\phi l}^{(3)}[11]\\
    &-0.358 C_{\phi l}^{(3)}[22]+0.0258 C_{\phi l}^{(3)}[ii]\bigg) \times 10^{-6} \ \textrm{GeV}
\end{split}
\end{align}
\begin{align}
\begin{split}
    \delta\Gamma&_\textrm{NLO}(\hsm\to Z\ell_i^+\ell_i^-) = 0.0416 \times 10^{-6} \ \textrm{GeV}\\
    &+\left(\frac{1 \ \textrm{TeV}}{\Lambda} \right)^2\bigg(-0.00726 C_{\phi} +0.18 C_{\phi B}+0.0218 C_{\phi \square}-0.0344 C_{\phi D}-0.185 C_{\phi W}\\
    &+0.0899 C_{\phi WB}+0.00287 C_{W}-0.00488 C_{eu}[ii33]-0.00789 C_{ll}[1122]\\
    &+0.0101 C_{ll}[1221]-0.00632 C_{lq}^{(1)}[ii33]+0.00595 C_{lq}^{(3)}[ii33]+0.00606 C_{lu}[ii33]\\
    &+0.00191 C_{\phi e}[ii]-0.00758 C_{\phi l}^{(1)}[ii]-0.0164 C_{\phi l}^{(3)}[11]-0.0164 C_{\phi l}^{(3)}[22]\\
    &-0.00512 C_{\phi l}^{(3)}[ii]+0.017 C_{\phi q}^{(1)}[33]-0.00108 C_{\phi q}^{(3)}[11]-0.00108 C_{\phi q}^{(3)}[22]\\
    &-0.0214 C_{\phi q}^{(3)}[33]-0.0191 C_{\phi u}[33]+0.00509 C_{qe}[33ii]+0.00243 C_{u\phi }[33]\\
    &-0.00655 C_{uW}[33]\bigg) \times 10^{-6} \ \textrm{GeV}\, ,
\end{split}
\end{align}
where we have dropped contributions smaller than $10^{-9}$ GeV in both $\Gamma_\mathrm{LO}$ and $\delta \Gamma_\mathrm{NLO}$.

Results for differential distributions are given in Sec.~\ref{sec:diff_dists}.
As pointed out in~\cite{Dawson:2024pft}, care should be taken when applying these results to LHC studies, since the experimental cut on $m_{\ell\ell}$ significantly impacts the results for some coefficients.

\subsubsection{\texorpdfstring{$\hsm\to Z \nu \bar{\nu}$}{H to Zvv}}

The process $\hsm\to Z\nu\bar{\nu}$ is the simplest three-body decay since there are no charged particles in the final state. At this order, there are no IR divergences or real emission contributions to consider.
We sum over final state neutrino flavours since they are unobservable, finding the result
\begin{align}
\begin{split}
    \Gamma_\textrm{LO}&(\hsm\to Z\nu\bar{\nu}) = 17.5 \times 10^{-6} \ \textrm{GeV} \\
    & + \left(\frac{1 \ \textrm{TeV}}{\Lambda}\right)^2\bigg(-0.231 C_{\phi B}+2.13 C_{\phi \square}-0.806 C_{\phi W}-0.433 C_{\phi WB}\\
    &+2.12 C_{ll}[1221]-0.0464 C_{\phi l}^{(1)}[11]-0.0464 C_{\phi l}^{(1)}[22]-0.0464 C_{\phi l}^{(1)}[33]\\
    &-2.08 C_{\phi l}^{(3)}[11]-2.08 C_{\phi l}^{(3)}[22]+0.0462 C_{\phi l}^{(3)}[33]-0.00114 C_{\phi q}^{(3)}[11]\\
    &-0.00114 C_{\phi q}^{(3)}[22]\bigg)  \times 10^{-6} \ \textrm{GeV}
\end{split}
\end{align}
\begin{align}
\begin{split}
    \delta\Gamma&_\textrm{NLO}(\hsm\to Z\nu\bar{\nu}) = 0.404 \times 10^{-6} \ \textrm{GeV}\\
    &+\left(\frac{1 \ \textrm{TeV}}{\Lambda}\right)^2 \bigg(-0.0431 C_{\phi} -0.0111 C_{\phi B}+0.148 C_{\phi \square}+0.0449 C_{\phi D}\\
    &-0.0209 C_{\phi W}+0.0172 C_{\phi WB}+0.00698 C_{W}-0.047 C_{ll}[1122]\\
    &+0.0881 C_{ll}[1221]+0.0114 C_{lq}^{(1)}[1133]+0.0114 C_{lq}^{(1)}[2233]+0.0114 C_{lq}^{(1)}[3333]\\
    &+0.0165 C_{lq}^{(3)}[1133]+0.0165 C_{lq}^{(3)}[2233]+0.0107 C_{lq}^{(3)}[3333]-0.0109 C_{lu}[1133]\\
    &-0.0109 C_{lu}[2233]-0.0109 C_{lu}[3333]+0.012 C_{\phi l}^{(1)}[11]+0.012 C_{\phi l}^{(1)}[22]\\
    &+0.0142 C_{\phi l}^{(1)}[33]-0.136 C_{\phi l}^{(3)}[11]-0.136 C_{\phi l}^{(3)}[22]-0.0121 C_{\phi l}^{(3)}[33]\\
    &+0.0924 C_{\phi q}^{(1)}[33]-0.00719 C_{\phi q}^{(3)}[11]-0.00719 C_{\phi q}^{(3)}[22]-0.112 C_{\phi q}^{(3)}[33]\\
    &-0.00138 C_{\phi u}[11]-0.00138 C_{\phi u}[22]-0.0964 C_{\phi u}[33]-0.00476 C_{uB}[33]\\
    &+0.015 C_{u\phi }[33]+0.00889 C_{uW}[33] \bigg)  \times 10^{-6} \ \textrm{GeV}\, ,
\end{split}
\end{align}
where we have dropped contributions smaller than $10^{-9}$ GeV. 

\subsubsection{\texorpdfstring{$\hsm\to Z q\bar{q}$}{H to Zqq}}

The process $\hsm\to Z q\bar{q}$ is similar to $Z\ell^+\ell^-$, with the addition of QCD real and virtual corrections. 
We consider only inclusive observables since these are of primary interest, although non-inclusive observables may be computed by using fragmentation functions as described in~\cite{Denner:2019vbn}.

At LO, the partial width for the $\hsm\to Zd_i \bar{d}_i$ process is the same for  $b\bar{b}$, $d\bar{d}$, and $s\bar{s}$ with appropriate flavour index swapping. Denoting by $i$ the flavour index of the down-type quark $d_i \in (d,s,b)$, we find a result of
\begin{align}
\begin{split}
    \Gamma_\textrm{LO}&(\hsm\to Zd_i\bar{d}_i) = 13.1 \times 10^{-6} \ \textrm{GeV} \\
    & + \left(\frac{1 \ \textrm{TeV}}{\Lambda}\right)^2\bigg(-1.58 C_{\phi B}+1.59 C_{\phi \square}+0.387 C_{\phi D}+0.801 C_{\phi W}-0.842 C_{\phi WB}\\
    &+1.59 C_{ll}[1221]-0.0207 C_{\phi d}[ii]-1.59 C_{\phi l}^{(3)}[11]-1.59 C_{\phi l}^{(3)}[22]+0.118 C_{\phi q}^{(1)}[ii]\\
    &+0.118 C_{\phi q}^{(3)}[ii]\bigg)  \times 10^{-6} \ \textrm{GeV}\, .
\end{split}
\end{align}

At NLO, the process $\hsm\to Zb\bar{b}$ is  special since there are more top-quark contributions with a diagonal CKM than for the light fermions. 
The $\hsm\to Zb\bar{b}$ result is given by
\begin{align}
\begin{split}
    \delta\Gamma&_\textrm{NLO}(\hsm\to Zb\bar{b}) = 0.047 \times 10^{-6} \ \textrm{GeV}\\
    &+\left(\frac{1 \ \textrm{TeV}}{\Lambda}\right)^2 \bigg( -0.0322 C_{\phi} -0.0113 C_{\phi B}+0.0795 C_{\phi \square}+0.0496 C_{\phi D}\\
    &-0.0742 C_{\phi G}+0.0261 C_{\phi W}+0.0934 C_{\phi WB}+0.00575 C_{W}-0.0349 C_{ll}[1122]\\
    &-0.0103 C_{ll}[1221]+0.00381 C_{lq}^{(3)}[1133]+0.00381 C_{lq}^{(3)}[2233]\\
    &+0.00137 C_{\phi d}[11]+0.00137 C_{\phi d}[22]+0.00201 C_{\phi d}[33]+0.00137 C_{\phi e}[11]\\
    &+0.00137 C_{\phi e}[22]+0.00137 C_{\phi e}[33]+0.00137 C_{\phi l}^{(1)}[33]-0.0173 C_{\phi l}^{(3)}[11]\\
    &-0.0175 C_{\phi l}^{(3)}[22]-0.0015 C_{\phi l}^{(3)}[33]-0.00137 C_{\phi q}^{(1)}[11]-0.00137 C_{\phi q}^{(1)}[22]\\
    &+0.0478 C_{\phi q}^{(1)}[33]-0.00451 C_{\phi q}^{(3)}[11]-0.00451 C_{\phi q}^{(3)}[22]-0.0779 C_{\phi q}^{(3)}[33]\\
    &-0.00274 C_{\phi u}[11]-0.00274 C_{\phi u}[22]-0.0842 C_{\phi u}[33]+0.00524 C_{qd}^{(1)}[3333]\\
    &-0.0587 C_{qq}^{(1)}[3333]-0.00311 C_{qq}^{(3)}[1133]-0.00311 C_{qq}^{(3)}[2233]+0.0195 C_{qq}^{(3)}[3333]\\
    &+0.0279 C_{qu}^{(1)}[3333]+0.00463 C_{uB}[33]-0.00488 C_{ud}^{(1)}[3333]-0.0055 C_{u\phi }[33]\\
    &-0.0452 C_{uW}[33]\bigg)  \times 10^{-6} \ \textrm{GeV}\, .
\end{split}
\end{align}

For the $\hsm\to Zd\bar{d}$ and $\hsm\to Zs\bar{s}$ channels, we have
\begin{align}
\begin{split}
    \delta\Gamma&_\textrm{NLO}(\hsm\to Zd_i\bar{d}_i)_{d_i \neq b} = 0.493 \times 10^{-6} \ \textrm{GeV}\\
    &+\left(\frac{1 \ \textrm{TeV}}{\Lambda}\right)^2 \bigg( -0.0322 C_{\phi }-0.0447 C_{\phi B}+0.134 C_{\phi \square}+0.0625 C_{\phi D}\\
    &-0.0741 C_{\phi G}+0.00995 C_{\phi W}+0.0801 C_{\phi WB}+0.0162 C_{W}-0.0349 C_{ll}[1122]\\
    &+0.0709 C_{ll}[1221]+0.00432 C_{lq}^{(3)}[1133]+0.00432 C_{lq}^{(3)}[2233]+0.00064 C_{\phi d}[ii]\\
    &+0.00137 C_{\phi d}[11]+0.00137 C_{\phi d}[22]+0.00137 C_{\phi d}[33]+0.00137 C_{\phi e}[11]\\
    &+0.00137 C_{\phi e}[22]+0.00137 C_{\phi e}[33]+0.00137 C_{\phi l}^{(1)}[33]-0.0988 C_{\phi l}^{(3)}[11]\\
    &-0.0988 C_{\phi l}^{(3)}[22]-0.0015 C_{\phi l}^{(3)}[33]-0.0293 C_{\phi q}^{(1)}[ii]-0.00137 C_{\phi q}^{(1)}[11]\\
    &-0.00137 C_{\phi q}^{(1)}[22]+0.0782 C_{\phi q}^{(1)}[33]-0.0182 C_{\phi q}^{(3)}[ii]-0.00451 C_{\phi q}^{(3)}[11]\\
    &-0.00451 C_{\phi q}^{(3)}[22]-0.1 C_{\phi q}^{(3)}[33]-0.00274 C_{\phi u}[11]-0.00274 C_{\phi u}[22]\\
    &-0.0906 C_{\phi u}[33]+0.00509 C_{qd}^{(1)}[33ii]-0.0583 C_{qq}^{(1)}[ii33]-0.00281 C_{qq}^{(3)}[iiii]\\
    &-0.00311 C_{qq}^{(3)}[1122]+0.0549 C_{qq}^{(3)}[ii33]-0.0354 C_{qq}^{(3)}[i33i]+0.0279 C_{qu}^{(1)}[ii33]\\
    &+0.00632 C_{uB}[33]-0.00488 C_{ud}^{(1)}[3311]+0.0105 C_{u\phi }[33]\\
    &-0.0456 C_{uW}[33]\bigg)  \times 10^{-6} \ \textrm{GeV}\,.
\end{split}
\end{align}

The up-type channels have the LO and NLO results,
\begin{align}
\begin{split}
    \Gamma_\textrm{LO}(\hsm\to& Zu_i\bar{u}_i) = 10.2 \times 10^{-6} \ \textrm{GeV} \\
    & + \left(\frac{1 \ \textrm{TeV}}{\Lambda}\right)^2\bigg(-1.75 C_{\phi B}+1.24 C_{\phi \square} +0.446 C_{\phi D}+1.15 C_{\phi W}\\
    &-0.852 C_{\phi WB}+1.24 C_{ll}[1221]-1.24 C_{\phi l}^{(3)}[11]-1.24 C_{\phi l}^{(3)}[22]\\
    &-0.0981 C_{\phi q}^{(1)}[ii]+0.0974 C_{\phi q}^{(3)}[ii]+0.0413 C_{\phi u}[ii]\bigg)  \times 10^{-6} \ \textrm{GeV}
\end{split}
\end{align}
\begin{align}
\begin{split}
    \delta\Gamma&_\textrm{NLO}(\hsm\to Zu_i\bar{u}_i) = 0.293 \times 10^{-6} \ \textrm{GeV}\\
    &+\left(\frac{1 \ \textrm{TeV}}{\Lambda}\right)^2 \bigg(-0.0251 C_{\phi }+0.131 C_{\phi B}+0.0931 C_{\phi \square}+0.0158 C_{\phi D}\\
    &-0.0578 C_{\phi G}-0.154 C_{\phi W}+0.164 C_{\phi WB}+0.0167 C_{W}-0.0272 C_{ll}[1122]\\
    &+0.038 C_{ll}[1221]+0.00337 C_{lq}^{(3)}[1133]+0.00337 C_{lq}^{(3)}[2233]+0.00139 C_{\phi d}[11]\\
    &+0.00139 C_{\phi d}[22]+0.00139 C_{\phi d}[33]+0.00139 C_{\phi e}[11]+0.00139 C_{\phi e}[22]\\
    &+0.00139 C_{\phi e}[33]+0.00139 C_{\phi l}^{(1)}[33]-0.0596 C_{\phi l}^{(3)}[11]-0.0596 C_{\phi l}^{(3)}[22]\\
    &-0.00106 C_{\phi l}^{(3)}[33]+0.02449 C_{\phi q}^{(1)}[ii]-0.00139 C_{\phi q}^{(1)}[11]-0.00139 C_{\phi q}^{(1)}[22]\\
    &+0.0643 C_{\phi q}^{(1)}[33]-0.01401 C_{\phi q}^{(3)}[ii]-0.00319 C_{\phi q}^{(3)}[11]-0.00319 C_{\phi q}^{(3)}[22]\\
    &-0.0842 C_{\phi q}^{(3)}[33]-0.00143 C_{\phi u}[ii]-0.00277 C_{\phi u}[11]-0.00277 C_{\phi u}[22]\\
    &-0.0776 C_{\phi u}[33]+0.0481 C_{qq}^{(1)}[ii33]+0.0144 C_{qq}^{(1)}[i33i]\\
    &-0.00213 C_{qq}^{(3)}[iiii]-0.00256 C_{qq}^{(3)}[1122]+0.0453 C_{qq}^{(3)}[ii33]+0.0152 C_{qq}^{(3)}[i33i]\\
    &-0.0231 C_{qu}^{(1)}[ii33]-0.0102 C_{qu}^{(1)}[33ii]+0.00862 C_{uB}[33]\\
    &+0.00794 C_{u\phi}[33]+0.0195 C_{uu}[ii33]+0.006 C_{uu}[i33i]\\
    &-0.0552 C_{uW}[33]\bigg)  \times 10^{-6} \ \textrm{GeV}\, ,
\end{split}
\end{align}

where $u_i \in (u,c)$ is the final state up-type quark flavour.
In all these channels we have dropped contributions smaller than $10^{-9}$ GeV in both $\Gamma_\mathrm{LO}$ and $\delta \Gamma_\mathrm{NLO}$.

\subsection{\texorpdfstring{$\hsm\to W f {\overline{f}^\prime}$}{H to Wff}}
\label{sec:wffsec}

\begin{figure}
\centering
\parbox{110pt}{\begin{fmffile}{ffW_re}
  \begin{fmfgraph*}(100,120)
    \fmfstraight
    \fmfright{o4,o3,o2,o1}
    \fmfleft{i2,h,i1}
    \fmftop{top}
    \fmfbottom{bot}
    \fmf{dashes,tension=1.6}{h,t1}
    \fmf{boson,tension=1.3}{t1,t3}
    \fmf{boson,tension=1.3}{t1,v1}
    \fmf{phantom,tension=1}{v1,top}
    \fmf{phantom,tension=1}{t3,bot}
    \fmfshift{16 down}{o1}
    \fmfshift{ 8 down}{o2}
    \fmfshift{ 8 up}{o3}
    \fmfshift{16 up}{o4}
    \fmf{fermion,tension=1.8}{o1,v1,o2}
    \fmf{boson,tension=1.8}{o4,t3,o3}
    \fmfv{l=$\bar{f}'$}{o1}
    \fmfv{l=$f$}{o2}
    \fmfv{l=$\gamma$}{o3}
    \fmfv{l=$h$}{h}
    \fmfv{l=$W^\pm$}{o4}
    \fmfblob{8}{t1} 
    \fmfv{l.a=120,l=\color{black}{$C_{\phi \square}$}}{t1}
  \end{fmfgraph*}
\end{fmffile}}
\qquad
\parbox{110pt}{
\begin{fmffile}{nlo_cuphi}
  \begin{fmfgraph*}(100,90)
    \fmfstraight
    \fmfright{o4,o3,o2,o1}
    \fmfleft{i2,h,i1}
    \fmftop{top}
    \fmfbottom{bot}
    \fmf{dashes,tension=1}{h,t1}
     \fmf{fermion,tension=.6}{t1,t3,t2,t1}
    \fmf{phantom,tension=.8}{i1,t2}
    \fmf{phantom,tension=.8}{t3,i2}
    \fmfshift{8 down}{o1}
    \fmfshift{ 4 down}{o2}
    \fmfshift{ 4 up}{o3}
    \fmfshift{8 up}{o4}
    \fmf{fermion,tension=.0001}{o1,v1,o2}
    \fmf{phantom,tension=1}{o1,t2,o2}
    \fmf{phantom,tension=1}{o3,t3}
    \fmf{boson,tension=1}{o4,t3}
    \fmf{boson,tension=.0001}{t2,v1}
    \fmfblob{8}{t1}
    \fmfv{l.d=12,l.a=110,l=\color{black}{$C_{u\phi}[33]$}}{t1}
    \fmfv{l=$h$}{h}
    \fmfv{l=$\bar{f}'$}{o1}
    \fmfv{l=$f$}{o2}
    \fmfv{l=$W^\pm$,l.a=-20}{o4}
  \end{fmfgraph*}
\end{fmffile}}
\qquad
\parbox{110pt}{\begin{fmffile}{nlo_Wqq_box}
    \begin{fmfgraph*}(100,80)
    \fmfleft{i1,i2,i3}
    \fmfright{o1,o2,o3}
    \fmftop{t1,t2}
    \fmfbottom{b1,b2}
    \fmf{dashes,t=1}{i2,v1} 
    \fmf{phantom,t=1}{v1,v2}
    \fmf{boson,t=1.1}{v2,o2}
    \fmffreeze
    \fmf{phantom,t=0}{vtl,i3} 
    \fmf{phantom,t=.5}{vtl,vtr} 
    \fmf{phantom,t=1}{vtr,o3} 
    \fmf{phantom,t=1}{v1,vtl,vtr} 
    \fmf{fermion,t=0}{vtr,v2} 
    \fmf{phantom,t=0}{vbl,i1} 
    \fmf{phantom,t=.5}{vbl,vbr} 
    \fmf{phantom,t=1}{vbr,o1} 
    \fmf{phantom,t=1}{v1,vbl,vbr} 
    \fmf{fermion,t=0,l.d=3}{v2,vbr}
    \fmffreeze
    \fmf{gluon,t=1,l.d=3}{v1,vbr}
    \fmf{gluon,t=1,l.d=3}{vtr,v1} 
    \fmfblob{8}{v1}
    \fmf{fermion,t=0}{vbr,o1}
    \fmf{fermion,t=0}{o3,vtr}
    \fmfv{l.d=7,l.a=115,l=\color{black}{$C_{\phi G}$}}{v1}
    \fmfv{l=$h$}{i2}
    \fmfv{l=$W^\pm$}{o2}
    \fmfv{l=$q$}{o1}
    \fmfv{l=$\bar{q}'$}{o3}
  \end{fmfgraph*}
\end{fmffile} }

    \caption{Representative diagrams for the $\hsm\to f\bar{f}'W^\pm$ decay process in the SMEFT. Dashed blobs represent SMEFT operator insertions.}
    \label{fig:h_to_ffW}
\end{figure}
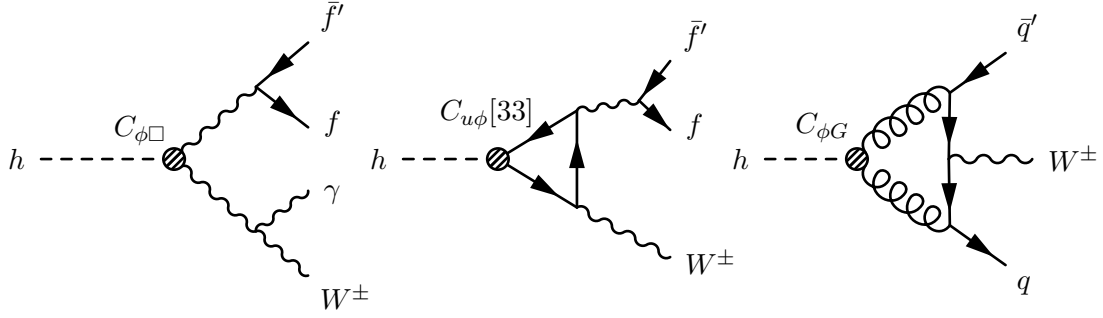
Just as for $\hsm\to Zf\bar{f}$, the SM corrections to $\hsm\to Wf\bar{f}'$ are a subset of the full $\hsm\to 4f$ process, which are known to NLO EW and QCD in the SM~\cite{Boselli:2015aha,Kaur:2023eyv,Bredenstein:2006rh,Bredenstein:2006ha}. 
The corrections most relevant for $\hsm\to Wf\bar{f}'$ are implemented in {\sc Prophecy4f}~\cite{Denner:2019fcr} and {\sc HDECAY}~\cite{Djouadi:2018xqq}.

In the SMEFT, we compute for the first time the physical $\hsm\to Wf\bar{f}'$ process at NLO, superseding the known (kinematically unphysical) $\hsm\to W W^*$ results~\cite{Dawson:2018liq}.
In our numerical results we have summed over $\hsm\to W^+ f\bar{f}'$ and the charge conjugate process $\hsm\to W^- f'\bar{f}$.

Since the $W$ is charged, real photon emission from the $W$ makes massive dipole subtraction necessary, as described in Sec.~\ref{sec:real_emission}.
The structure of the QED IR poles for $\hsm\to Wf\bar{f}'$ is the same as for $q{\overline{q}}\rightarrow Wj$~\cite{Kuhn:2007cv}, with appropriate swapping of initial and final state legs.
We use the expressions for all dipole functions and phase space mappings given in~\cite{Catani:2002hc} for the $W$ as the emitter or spectator, and use the massless formulas given in~\cite{Denner:2019vbn} for the QCD and QED pairing of the fermion legs since we consider only the top quark to be massive.

\subsubsection{\texorpdfstring{$\hsm\to W^\pm \ell^\mp \nu$}{H to Wlv}}

We first consider the simpler leptonic process $\hsm\to W^\pm \ell_i^\mp \nu_i$, where $\ell_i = (e,\mu,\tau)$. 
There are no QCD corrections to this process at one-loop, and the QED dipole subtraction has only two terms, with the $W$ and $\ell$ each as emitter/spectator. 
We find at LO (summing over $W^\pm$, but not over $i$),
\begin{align}
\begin{split}
    \Gamma_\textrm{LO}&(\hsm\to W^\pm \ell_i \nu_i) = 87.7 \times 10^{-6} \ \textrm{GeV}\\
    &+\left(\frac{1 \ \textrm{TeV}}{\Lambda}\right)^2 \bigg(10.6 C_{\phi \square}-2.66 C_{\phi D}-7.92 C_{\phi W}+10.6 C_{ll}[1221] 
    \\ &-10.6 C_{\phi l}^{(3)}[11]-10.6 C_{\phi l}^{(3)}[22]+1.63 C_{\phi l}^{(3)}[ii] -0.00693 C_{\phi q}^{(3)}[11] \\
    & - 0.00693 C_{\phi q}^{(3)}[22]\bigg) \times 10^{-6} \ \textrm{GeV}\, ,
\end{split}
\end{align}
where only the coefficient $C_{\phi l}^{(3)}[ii]$ depends on the lepton flavour $i$.
Here we have truncated coefficients arising from the width expansion that contribute less than $10^{-9}$ GeV.

At NLO, we find
\begin{align}
\begin{split}
    \delta\Gamma&_\textrm{NLO}(\hsm\to W^\pm\ell^\mp_i\nu) = 2.99 \times 10^{-6} \ \textrm{GeV}\\
    &+\left(\frac{1 \ \textrm{TeV}}{\Lambda}\right)^2 \bigg(-0.173 C_{\phi} -0.0291 C_{\phi B}+0.834 C_{\phi \square}-0.334 C_{\phi D}-0.392 C_{\phi W}\\
    &-0.284 C_{\phi WB}+0.0283 C_{W}+0.385 C_{ll}[1221]-0.00477 C_{ll}[1ii1]\\
    &-0.00477 C_{ll}[2ii2]-0.00477 C_{ll}[3ii3]+0.029 C_{lq}^{(3)}[1133]+0.029 C_{lq}^{(3)}[2233]\\
    &-0.0187 C_{lq}^{(3)}[ii11]-0.0187 C_{lq}^{(3)}[ii22]+0.15 C_{lq}^{(3)}[ii33]-0.0109 C_{\phi l}^{(1)}[11]\\
    &-0.0109 C_{\phi l}^{(1)}[22]+0.00362 C_{\phi l}^{(1)}[ii]-0.811 C_{\phi l}^{(3)}[11]-0.811 C_{\phi l}^{(3)}[22]\\
    &-0.0604 C_{\phi l}^{(3)}[ii]-0.0448 C_{\phi q}^{(3)}[11]-0.0448 C_{\phi q}^{(3)}[22]-0.445 C_{\phi q}^{(3)}[33]\\
    &-0.0291 C_{u\phi }[33]+0.118 C_{uW}[33]\bigg) \times 10^{-6} \ \textrm{GeV}\, ,
\end{split}
\end{align}
where we have not truncated any coefficients, and only $C_{ll}$, $C_{lq}^{(3)}$, $C_{\phi l}^{(1)}$, and $C_{\phi l}^{(3)}$ have coefficients that depend on the final state lepton flavour $i$.

\subsubsection{\texorpdfstring{$\hsm\to W^\pm q{\bar {q}}'$}{H to Wqq}}

We now move on to the hadronic process $\hsm\to W^\pm q_i {\bar{q}}_i'$, where $q_i{\bar{q}}_i' = (ud,cs)$.
Since we have final state quarks we must consider QCD corrections as well.
The QCD dipole subtraction is identical to the $\hsm\to Zq\bar{q}$ case, whereas the QED dipole subtraction consists of six terms -- two for each emitter/spectator pair among $W^\pm q {\bar{q}}'$.
Summing over $W^\pm$ but not over the flavour index $i$, we find the result
\begin{align}
\begin{split}
    \delta\Gamma&_\textrm{LO}(\hsm\to W^\pm q_i {\bar{q}}'_i) =  263\times 10^{-6} \ \textrm{GeV}\\
    &+\left(\frac{1 \ \textrm{TeV}}{\Lambda}\right)^2 \bigg(31.9 C_{\phi\square}-7.97 C_{\phi D}-23.8 C_{\phi W}+31.9 C_{ll}[1221]\\
    &-31.9 C_{\phi l}^{(3)}[11]-31.9 C_{\phi l}^{(3)}[22]+4.88 C_{\phi q}^{(3)}[ii]-0.0208 C_{\phi q}^{(3)}[11]\\
    &-0.0208 C_{\phi q}^{(3)}[22] -0.00693 C_{\phi l}^{(3)}[33]\bigg) \times 10^{-6} \ \textrm{GeV}\, ,
\end{split}
\end{align}
where we have once again truncated coefficients coming from the expansion of $\Gamma_W$ that contribute less than $10^{-9}$ GeV.

At NLO, we find a correction
\begin{align}
\begin{split}
    \delta\Gamma&_\textrm{NLO}(\hsm\to W^\pm q_i {\overline{q}}'_i) =  18.8\times 10^{-6}  \ \textrm{GeV}\\
    &+\left(\frac{1 \ \textrm{TeV}}{\Lambda}\right)^2 \bigg(-0.519 C_{\phi }-0.0728 C_{\phi B}+3.7 C_{\phi \square}-1.3 C_{\phi D}\\
    &-1.34 C_{\phi G}-4.48 C_{\phi W}-0.839 C_{\phi WB}+0.0848 C_{W}+2.35 C_{ll}[1221]\\
    &-0.0187 C_{lq}^{(3)}[11ii]+0.0869 C_{lq}^{(3)}[1133]-0.0187 C_{lq}^{(3)}[22ii]+0.0869 C_{lq}^{(3)}[2233]\\
    &-0.0187 C_{lq}^{(3)}[33ii]-0.0326 C_{\phi l}^{(1)}[11]-0.0326 C_{\phi l}^{(1)}[22]-3.63 C_{\phi l}^{(3)}[11]\\
    &-3.63 C_{\phi l}^{(3)}[22]-0.0448 C_{\phi l}^{(3)}[33]-0.00357 C_{\phi q}^{(1)}[ii]+0.0467 C_{\phi q}^{(3)}[ii]\\
    &-0.134 C_{\phi q}^{(3)}[11]-0.134 C_{\phi q}^{(3)}[22]-1.33 C_{\phi q}^{(3)}[33]-0.0143 C_{qq}^{(1)}[iiii]\\
    &-0.0143 C_{qq}^{(1)}[1221]+0.124 C_{qq}^{(1)}[i33i]-0.0979 C_{qq}^{(3)}[iiii]\\
    &-0.112 C_{qq}^{(3)}[1122]+0.901 C_{qq}^{(3)}[ii33]+0.0143 C_{qq}^{(3)}[1221]\\
    &-0.124 C_{qq}^{(3)}[i33i]-0.0872 C_{u\phi }[33]+0.354 C_{uW}[33]\bigg) \times 10^{-6} \ \textrm{GeV}\, .
\end{split}
\end{align}

\subsection{\texorpdfstring{$\hsm\to f\bar{f}$}{H to ff}}
\label{sec:hffsection}

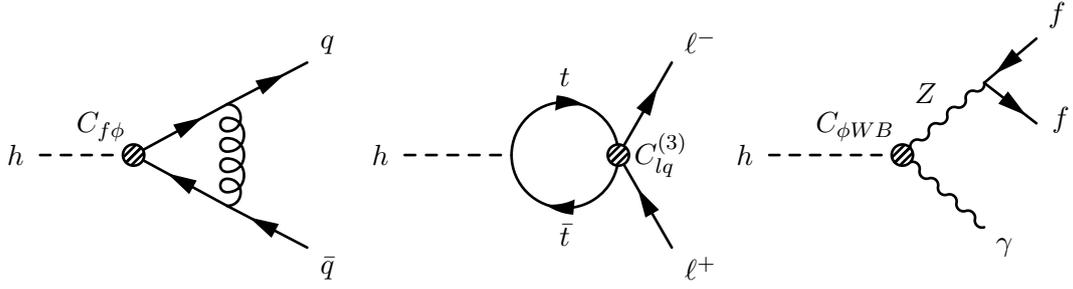
\begin{figure}
\centering
\begin{fmffile}{ff_virt_1}
  \begin{fmfgraph*}(100,120)
    \fmfstraight
    \fmfleft{i2,g,i1}
    \fmfright{o4,o3,o2}
    \fmftop{top}
    \fmfbottom{bot}
    \fmf{dashes,tension=2}{g,h}
    \fmf{fermion,tension=1,l.s=right}{h,v1}
    \fmf{fermion,tension=1,l.s=right}{v2,h}
    \fmf{phantom,tension=.5}{top,v1}
    \fmf{phantom,tension=.5}{v2,bot}
    \fmfshift{25 down}{o2}
    \fmfshift{25 up}{o4}
    \fmf{fermion,tension=1.5}{o4,v2}
    \fmf{gluon,tension=.65}{v2,v1}
    \fmf{fermion,tension=1.5}{v1,o2}
   \fmfblob{0.08w}{h}
    \fmfv{l=$C_{f\phi}$,l.a=120}{h}
    \fmfv{l=$h$}{g}
    \fmfv{l=$\bar{q}$}{o4}
    \fmfv{l=$q$}{o2}
  \end{fmfgraph*}
\end{fmffile} 
\qquad
\begin{fmffile}{ff_virt_2}
  \begin{fmfgraph*}(100,120)
    \fmfstraight
    \fmfleft{i2,g,i1}
    \fmfright{o4,o3,o2}
    \fmftop{top}
    \fmfbottom{bot}
    \fmf{dashes,tension=1}{g,h}
    \fmf{fermion,tension=.5,left=1,label=$t$}{h,v1}
    \fmf{fermion,tension=.5,left=1,label=$\bar{t}$}{v1,h}
      \fmfshift{25 down}{o2}
  
      \fmfshift{25 up}{o4}
    \fmf{fermion,tension=1}{o4,v1}
    \fmf{fermion,tension=1}{v1,o2}
    \fmfblob{0.08w}{v1}
    \fmfv{l=$C_{lq}^{(3)}$,l.a=0}{v1}
    \fmfv{l=$h$}{g}
    \fmfv{l=$\ell^+$}{o4}
    \fmfv{l=$\ell^-$}{o2}
  \end{fmfgraph*}
\end{fmffile} 
\qquad 
\begin{fmffile}{ffgamma}
  \begin{fmfgraph*}(100,120)
    \fmfstraight
    \fmfleft{i2,g,i1}
    \fmfright{o4,o3,o2,o1}
    \fmftop{top}
    \fmfbottom{bot}
    \fmf{dashes,tension=1.6}{g,h}
    \fmf{photon,tension=1.3,label=$Z$,l.s=right}{v1,h}
    \fmf{photon,tension=1.3,l.s=right}{h,v2}
    \fmf{phantom,tension=1.0}{top,v1}
    \fmf{phantom,tension=1.0}{v2,bot}
    \fmfshift{16 down}{o1}
    \fmfshift{ 8 down}{o2}
    \fmfshift{ 8 up}{o3}
    \fmfshift{16 up}{o4}
    \fmf{fermion,tension=1.8}{o1,v1,o2}
    \fmf{phantom,tension=1.8}{o4,v2,o3}
    \fmfblob{0.08w}{h}
    \fmfv{l=$C_{\phi WB}$,l.a=120}{h}
    \fmfv{l=$h$}{g}
    \fmfv{l=$\bar{f}$}{o1}
    \fmfv{l=$f$}{o2}
    \fmfv{l=$\gamma$}{v2}
  \end{fmfgraph*}
  \vspace{-.5cm}
\end{fmffile} 
    \caption{Representative diagrams for the $\hsm\to f\bar{f}$ decay process in the SMEFT at NLO. The dashed blobs represent SMEFT operator insertions.}
    \label{fig:h_to_ff}
\end{figure}

The Higgs decays to $f\bar{f}$ are the most important branching ratios, constituting nearly 70\% of the total Higgs width in the SM.
The decays $\hsm\rightarrow b\bar{b},c\bar{c}$ are known exactly at NNLO QCD in the SM~\cite{Bernreuther:2018ynm,Primo:2018zby,Behring:2019oci,Somogyi:2020mmk,Wang:2024ilc}, with further results available in various approximations~\cite{Herzog:2017dtz,Mondini:2019gid,Chen:2023fba,Wang:2023xud}.
The state of the art EW corrections in the SM~\cite{Chetyrkin:1996wr,Kwiatkowski:1994cu,Kniehl:1994ju,Kataev:1997cq,Djouadi:1997rj} for $\hsm\rightarrow f\bar{f}$ are one-loop exact and known at two-loops in the $m_t\rightarrow \infty$ expansion, as well two-loop mixed QCD-EW order~\cite{Mihaila:2015lwa}.
The most numerically relevant corrections are implemented in  {\sc MadGraph }~\cite{Alwall:2014hca} and {\sc HDECAY}~\cite{Djouadi:2018xqq}.

In the SMEFT, all relevant $\hsm\to f\bar{f}$ decays are known at NLO in the dimension-6 SMEFT~\cite{Gauld:2016kuu,Cullen:2019nnr,Cullen:2020zof}. 
We have implemented an independent calculation for all  $f \in (b,c,s,\tau,\mu)$.  For the $\hsm\rightarrow f\bar{f}$ processes, we must keep the fermion mass $m_f$ nonzero to have a non- vanishing SM contribution.
The LO widths are particularly simple since they are two-body decays:
\begin{align}
    \Gamma_\textrm{LO}(\hsm\to f_i\bar{f}_i) = \frac{\sqrt{\mhsm^2-4m_{f_i}^2}}{16\pi \mhsm^2}|\mathcal{A}^{(0)}(\hsm\to f_i\bar{f}_i)|^2
\end{align}
where the amplitude squared to linear order in the dimension-6 SMEFT coefficients is given by (omitting the colour sum for quarks)
\begin{align}
    |\mathcal{A}^{(0)}(\hsm\to f_i\bar{f}_i)|^2 =& |\mathcal{A}^{(0)}_\mathrm{SM}|^2 + \frac{2}{\Lambda^2} \mathrm{Re}(\mathcal{A}_\mathrm{SM}^{(0)*}\mathcal{A}_\mathrm{EFT}^{(0)}) \, ,
\end{align}
\begin{align}
    |\mathcal{A}^{(0)}_\mathrm{SM}|^2 &= 2\sqrt{2} G_\mu m_{f_i}^2 (\mhsm^2 -4m_{f_i}^2) \, ,
\end{align}
\begin{align}
\begin{split}
   \frac{2}{\Lambda^2} \mathrm{Re}(\mathcal{A}_\mathrm{SM}^{(0)*}\mathcal{A}_\mathrm{EFT}^{(0)}) =& \frac{m^2_{f_i}(\mhsm^2 -4m_{f_i}^2)}{\Lambda^2}\bigg[-\frac{2^{5/4}}{m_{f_i}\sqrt{G_\mu}}C_{f\phi}[ii] \\ 
   &+4C_{\phi\square}-C_{\phi D} -2C_{\phi l}^{(3)}[11]-2C_{\phi l}^{(3)}[22]+2C_{ll}[1221]\bigg] \, .
\end{split}
\end{align}

At NLO, each $\hsm\rightarrow f_i\bar{f_i}$ process must be considered separately.
Representative diagrams at NLO for these processes are shown in Fig.~\ref{fig:h_to_ff}.
For each process, we include the effects of nonzero $m_t,m_b,m_\tau$, as well as all masses greater than or equal to $m_{f_i}$ for consistency.
We perform on-shell renormalization for all fermion masses, and  discuss the effects of $\overline{\mathrm{MS}}$ masses at the end of this section. Since we consider $m_{f_i} \neq 0$, we use the massive dipole subtraction procedure discussed in Sec.~\ref{sec:real_emission} to handle IR singularities. 
As a cross check, we have also verified our subtraction procedure with a phase space slicing method~\cite{Denner:2019vbn} for these processes.

Since the $Z$ and $W$ do not appear as external states in this process, we consistently include their widths by using the complex mass scheme.
Including the width is important to regulate a divergence in the real emission contribution to $C_{\phi WB}$, $C_{\phi B}$, and $C_{\phi W}$ through the third diagram depicted in Fig.~\ref{fig:h_to_ff} when the $Z$ goes on-shell. 

We consider the two leptonic final states $\tau^+\tau^-$ and $\mu^+\mu^-$.
We do not consider the $e^+e^-$ channel since the electron mass suppresses the $\hsm \rightarrow e^+e^-$ rate to be unobservably small.
For $\hsm\rightarrow \tau^+\tau^-$, we keep $m_t$, $m_b$, and $m_\tau$ nonzero, while taking all lighter fermions massless. 
For $\hsm \rightarrow \mu^+\mu^-$, we take all of $m_t$, $m_b$, $m_\tau$, $m_c$, and $m_\mu$ to be nonzero, while taking lighter fermions massless.

We consider three hadronic final states: $b\bar{b}$, $c\bar{c}$, and $s\bar{s}$. 
The calculation is very similar to $\ell^+\ell^-$, with the addition of virtual and real emission QCD contributions.
For $\hsm\rightarrow b\bar{b}$, we take nonzero $m_t$, $m_b$, and $m_\tau$.
For $\hsm\rightarrow c\bar{c}$, we take nonzero $m_t$, $m_b$, $m_\tau$, and $m_c$.
Finally, we include results for $\hsm\to s\bar{s}$ for completeness, taking nonzero $m_t$, $m_b$, $m_\tau$, $m_c$, $m_\mu$, and $m_s$.

\subsubsection{\texorpdfstring{$\overline{\mathrm{MS}}$ masses in $\hsm\to q {\overline{q}}$}{H to qq}}
\label{sec:msbar}

For the channels $\hsm\rightarrow q\bar{q}$, it is well known that the choice of renormalization scheme for the quark masses $m_b$, $m_c$, and $m_s$ can make a significant numerical difference~\cite{LHCHiggsCrossSectionWorkingGroup:2016ypw}. Therefore, while our results are mainly obtained in terms of OS SM parameters, we have considered also the $\overline{\mathrm{MS}}$ scheme for the light quark masses. The work in Ref.~\cite{Gauld:2016kuu} and subsequent papers have a quite in-depth discussion on the relation between the two schemes for $\hsm\rightarrow q\bar{q}$ in the SMEFT. Here we summarize the main results presented in those papers, and refer to the original publications for more details.

In general, the relation between OS and $\overline{\mathrm{MS}}$ masses can be written as~\cite{Chetyrkin:2000yt,Herren:2017osy,Cullen:2020zof}
\begin{align}
    m_q=m_q(\mu)(1-\delta_q(\mu)),
\end{align}
where $m_q$ and $m_q(\mu)$ are the masses of a generic quark in the OS $\overline{\mathrm{MS}}$ and schemes, respectively. The function  $\delta_q(\mu)$ includes both NLO QCD and electroweak corrections in the SMEFT and in general depends on the $\overline{\mathrm{MS}}$ mass of the quark, although one can use the OS mass instead, as the difference is of higher order. 
To obtain the relation between the Higgs width calculated in the two schemes
\begin{align}
    \Gamma_{\overline{\mathrm{MS}}}(\hsm\rightarrow q\bar{q})=\Gamma_\mathrm{OS}(\hsm\rightarrow q\bar{q})+\Delta\Gamma(\hsm\rightarrow q\bar{q})
\end{align}
it is sufficient to replace the OS mass with its expression in terms of the $\overline{\mathrm{MS}}$ mass and then expand to linear order in $\delta_q(\mu)$. 

Finally, as the input values for the $\overline{\mathrm{MS}}$ quark masses are usually defined at or around the mass scale, we must evolve them to the scale $\mu = \mhsm$ before obtaining the final result for the partial widths.
This evolution depends on SMEFT coefficients, and so requires an implementation of the RGEs between $m_q$ and $\mu$.
This is beyond the current scope of {\sc NEWiSH}, and so we cannot implement these in full generality.
However, since this evolution is numerically important for these processes, we have evolved the Wilson coefficients appearing in $\delta_q(\mu)$ by using {\sc DsixTools}~\cite{Celis:2017hod,Fuentes-Martin:2020zaz}, which includes the full one-loop anomalous dimension matrix in SMEFT, with both QCD and EW corrections. The SM input masses are evolved with four-loop QCD beta functions in {\sc RunDec}~\cite{Chetyrkin:2000yt,Herren:2017osy}, and hard-coded the result for the specific inputs defined in Sec.~\ref{sec:inputs}.
Numerical results using these inputs are also in App.~\ref{sec:app_num}.

\subsection{\texorpdfstring{$\hsm\to VV'$}{H to VV'}}
\label{sec:htovv}

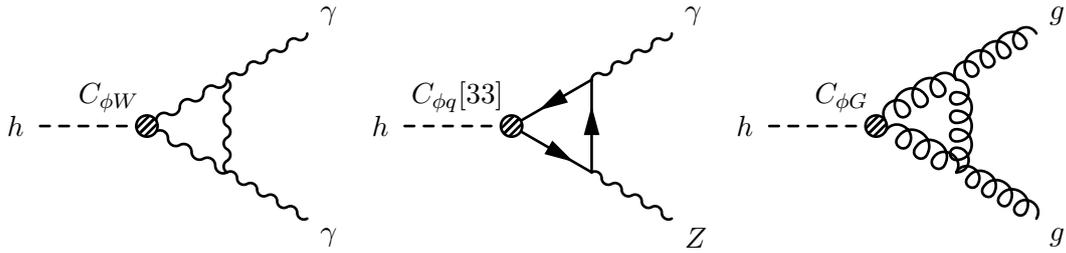
\begin{figure}
\centering
\begin{fmffile}{VV_1}
  \begin{fmfgraph*}(100,120)
    \fmfstraight
    \fmfleft{i2,g,i1}
    \fmfright{o4,o3,o2}
    \fmftop{top}
    \fmfbottom{bot}
    \fmf{dashes,tension=3}{g,h}
    \fmf{photon,tension=2,l.s=right}{o2,v1,h}
    \fmf{photon,tension=2,l.s=right}{o4,v2,h}
    \fmfshift{25 down}{o2}
    \fmfshift{25 up}{o4}
    \fmf{photon,tension=0}{v2,v1}
    \fmfblob{0.08w}{h}
    \fmfv{l=$C_{\phi W}$,l.a=120}{h}
    \fmfv{l=$h$}{g}
    \fmfv{l=$\gamma$}{o4}
    \fmfv{l=$\gamma$}{o2}
  \end{fmfgraph*}
\end{fmffile} 
\qquad
\begin{fmffile}{VV_2}
  \begin{fmfgraph*}(100,120)
    \fmfstraight
    \fmfleft{i2,g,i1}
    \fmfright{o4,o3,o2}
    \fmftop{top}
    \fmfbottom{bot}
    \fmf{dashes,tension=3}{g,h}
    \fmf{phantom,tension=2,l.s=right}{o2,v1,h}
    \fmf{phantom,tension=2,l.s=right}{o4,v2,h}
    \fmf{photon,tension=0,l.s=right}{o2,v1}
    \fmf{photon,tension=0,l.s=right}{o4,v2}
    \fmfshift{25 down}{o2}
    \fmfshift{25 up}{o4}
    \fmf{fermion,tension=0}{v2,v1,h,v2}
    \fmfblob{0.08w}{h}
    \fmfv{l=$C_{\phi q}[33]$,l.a=120}{h}
    \fmfv{l=$h$}{g}
    \fmfv{l=$Z$}{o4}
    \fmfv{l=$\gamma$}{o2}
  \end{fmfgraph*}
\end{fmffile} 
\qquad 
\begin{fmffile}{VV_3}
  \begin{fmfgraph*}(100,120)
    \fmfstraight
    \fmfleft{i2,g,i1}
    \fmfright{o4,o3,o2}
    \fmftop{top}
    \fmfbottom{bot}
    \fmf{dashes,tension=3}{g,h}
    \fmf{gluon,tension=2,l.s=right}{o2,v1,h}
    \fmf{gluon,tension=2,l.s=right}{o4,v2,h}
    \fmfshift{25 down}{o2}
    \fmfshift{25 up}{o4}
    \fmf{gluon,tension=0}{v2,v1}
    \fmfblob{0.08w}{h}
    \fmfv{l=$C_{\phi G}$,l.a=120}{h}
    \fmfv{l=$h$}{g}
    \fmfv{l=$g$}{o4}
    \fmfv{l=$g$}{o2}
  \end{fmfgraph*}
\end{fmffile} 
\vspace{-.5cm}
    \caption{Representative diagrams for the $\hsm\to VV'$ decay processes in the SMEFT at NLO. Dashed blobs represent SMEFT operator insertions.}
    \label{fig:h_to_VV}
\end{figure}

The processes $\hsm\rightarrow VV'$ with $VV'\in(gg,\gamma\gamma,Z\gamma)$ are unique since they first occur at the one-loop level in the SM, but have tree-level contributions in the SMEFT. 
Representative diagrams are shown in Fig.~\ref{fig:h_to_VV}.

For our implementation of these processes, we only include dimension-6, one-loop accurate results\footnote{The numerical effects of the partial higher order terms included in \cite{Martin:2023fad,Corbett:2021cil} are generally small, so we leave their implementation in {\sc NEWiSH} to future work.}.
To be explicit, the virtual amplitude for $\hsm \to VV'$ in the SMEFT is given by
\begin{equation}
    \mathcal{A}_{VV'} = \frac{1}{\Lambda^2}\mathcal{A}_{\text{EFT}}^{(0)}+ \frac{1}{(4\pi)^2}\mathcal{A}_\text{SM}^{(1)} + \frac{1}{(4\pi \Lambda)^2}\mathcal{A}_\text{EFT}^{(1)}  + \ldots
\end{equation}
We define LO and NLO widths given by 
\begin{equation}
    \Gamma_\textrm{LO}(\hsm\rightarrow VV') = \frac{c_{VV'}}{16\pi \mhsm}\left[\frac{2}{(4\pi \Lambda)^2} \mathrm{Re}\left(\mathcal{A}_\text{EFT}^{(0)*}\mathcal{A}_\text{SM}^{(1)} \right)\right] \, ,
\end{equation}
\begin{equation}
    \delta\Gamma_\textrm{Virt}(\hsm\rightarrow VV') = \frac{c_{VV'}}{16\pi \mhsm}\bigg[\frac{1}{(4\pi)^4}|\mathcal{A}_\text{SM}^{(1)}|^2 + \frac{2}{(4\pi)^4 \Lambda^2} \mathrm{Re}\left(\mathcal{A}_\text{EFT}^{(1)*}\mathcal{A}_\text{SM}^{(1)} \right)\bigg] \, ,
    \label{eq:virthvv}
\end{equation}
where the prefactor $c_{VV'}$ is equal to $1-M_Z^2/\mhsm^2$ for $\hsm\rightarrow Z\gamma $  and $1$ for $\hsm\rightarrow (\gamma\gamma,gg)$. We include a nonzero $m_b$ in the SM virtual amplitudes, but other light quarks are considered massless. 
For $\hsm\rightarrow \gamma\gamma, Z \gamma$, there are no real emission contributions so this is the full result up to one-loop order and
and $\Gamma_\mathrm{NLO}\equiv \Gamma_\mathrm{LO}+\delta \Gamma_\mathrm{Virt}$.

In the SM, the state of the art for $\hsm\to \gamma\gamma$ is NLO EW (two-loop)~\cite{Djouadi:1997rj,Actis:2008ts,Degrassi:2005mc,Passarino:2007fp}, full NNLO (three-loop) QCD~\cite{Maierhofer:2012vv,Niggetiedt:2020sbf}, and N$^3$LO QCD~\cite{Davies:2021zbx} in the $m_t\rightarrow \infty$ limit.
The $\hsm\rightarrow Z \gamma$ process is known in the SM at NLO EW~\cite{Chen:2024vyn,Sang:2024vqk} and NLO QCD~\cite{Gehrmann:2015dua,Bonciani:2015eua,SPIRA1992350}, which are both two-loop order.
In the SMEFT, $\hsm \to \gamma\gamma$ has been calculated in~\cite{Dawson:2018liq,Dedes:2018seb,Hartmann:2015aia} at one-loop using both the $(M_W,M_Z,G_\mu)$ and $(\alpha ,M_Z,G_\mu)$ input schemes.
The $\hsm \to Z\gamma$ process is likewise known at one-loop~\cite{Dawson:2018pyl,Dedes:2019bew}.

The SM result for $\hsm\to gg$ is a component of the full $\hsm\to \mathrm{hadrons}$ process, which is known  to approximate N$^4$LO QCD~\cite{Herzog:2017dtz,Chen:2023fba} (see also~\cite{Huss:2025nlt,Proceedings:2019vxr,Spira:2016ztx} for a more complete discussion).

For $\hsm \to gg$, there is in addition to the virtual correction of Eq. \eqref{eq:virthvv}, a real emission contribution proportional to $C_{\phi G}$ from $\hsm\rightarrow \Sigma_i f_i{\overline{f}_i} g$ and $\hsm\rightarrow ggg$.  These contributions are  trivially found from the $\hsm\rightarrow gg$ NLO SM calculation in the $m_t\rightarrow \infty$ limit~\cite{Spira:2016ztx}.
We note that in the NLO SMEFT, the SM contribution  occurs through the one-loop triangle graph and SM corrections to this diagram are higher order.  To  include higher order  SM corrections consistently, the two-loop contributions proportional to $C_{\phi G}$ would be needed, along with one-loop real contributions~\cite{Deutschmann:2017qum}.

For operators other than $C_{\phi G}$, the contributions to $\hsm\rightarrow gg$  first arise at one-loop and can be included to ${\cal{O}}(\frac{1}{16\pi^2\Lambda^4})$.  In order to keep the power counting consistent, we have chosen not to include these terms as they are numerically small~\cite {Asteriadis:2022ras}.  In addition, the decay $\hsm\rightarrow gg$ has been calculated to all orders in $\frac{v^2}{\Lambda^2}$ using geoSMEFT~\cite{Martin:2023fad,Corbett:2021cil}.

We present numerical results for  $\hsm \to VV'$, $VV' \in (\gamma\gamma,Z\gamma,gg)$ in App.~\ref{sec:apphvv}.

\subsection{\texorpdfstring{$\hsm\to ggZ$}{H to ggZ}}

The process $\hsm\to ggZ$ is a rare loop-induced decay in the SM~\cite{dEnterria:2025rjj,Abbasabadi:2008zz,Kniehl:1990yb}, with a very small rate.
In the SMEFT, it may be determined from the calculation of $gg\rightarrow Z\hsm$, which is known at NLO in the SMEFT~\cite{Rossia:2023hen}.
The process can be written as
\begin{equation}
    g^a_{\mu} (p_1)\, g^b_{\nu}(p_2) \rightarrow h(-p_3)\, Z (-p_4), 
\end{equation}
The Feynman amplitudes for these decay  scatterings are given by
\begin{align}
      \mathcal{A}_{hZ} & = i \sqrt{2} \, \frac{M_Z G_{\mu} \alpha_S(\mu_R)}{ \pi} \, \delta_{a b} \epsilon^a_{\mu}(p_1)  \epsilon^b_{\nu}(p_2)   \epsilon^*_{\rho}(p_3)   \hat{\mathcal{A}}^{\mu\nu }_{hZ}(p_1,p_2,p_3)
\end{align}
where $\epsilon(p_i)$ is the polarisation vector of the $i_{th}$ particle, $G_{\mu}$ is the Fermi constant and $\alpha_{S}$ is the QCD fine structure constant defined at the renormalization scale $\mu_R$. For  $\hsm Z$ production we have:
\begin{align}
      \mathcal{A}^{\mu \nu \rho}_{\hsm Z} & =\sum_i^7 \mathcal{P}^{\mu \nu \rho}_i {A_i}_{\hsm Z}\, ,
\end{align}
where $\mathcal{P}_i$  are a basis of orthonormal projectors and ${A_i}_{\hsm Z}$ are the associated scalar form factors~\cite{Bellafronte:2022jmo}.
These latter depend only on scalar quantities, namely the top quark mass  $m_t$, the masses of the external particles and
the partonic Mandelstam variables. 

Since the process is loop-induced in both the SM and the SMEFT, the partial width is negligibly small, though the momentum swapped process $gg\rightarrow Z\hsm $ is important at the LHC~\cite{Rossia:2023hen}.
For completeness, we find explicitly:
\begin{align}
\begin{split}
    \Gamma_\textrm{NLO}&(\hsm\to ggZ) =  2.26\times 10^{-9} \ \textrm{GeV} \\ 
    &+\left(\frac{1 \ \textrm{TeV}}{\Lambda}\right)^2\bigg(0.274 C_{\phi \square}+0.00711 C_{\phi G}+0.274 C_{ll}[1221]\\
    &-0.274 C_{\phi l}^{(3)}[11]-0.274 C_{\phi l}^{(3)}[22]+0.0365 C_{uG}[33]\bigg) \times 10^{-9} \ \textrm{GeV}\, ,
\end{split}
\end{align}
where we have dropped contributions smaller than $10^{-11}$ GeV, and there are no tree-level  or real emission contributions.

\subsection{\texorpdfstring{$\hsm\to 4f$}{H to 4f} and the narrow width approximation}\label{sec:h24f}
Thanks to our implementation of $\hsm \to f\bar{f}V$, we have access to the $\hsm \to 4 f$ processes at NLO through the narrow width approximation (NWA). 
At LO, the full results are known and have been studied in~\cite{Brivio:2019myy}.
We use the complex mass scheme at LO for the $4f$ process and neglect the $W$ and $Z$ widths in the NLO $\hsm \to f_1\bar{f}_2V$ corrections which are combined with the NLO $V\rightarrow f_3 {\overline{f}}_4$ contributions using the NWA.

For the total $\hsm \to 4f$ prediction, a subtlety appears in the treatment of the $W$ and $Z$ widths.
In the usual approach of the complex mass scheme, one takes the experimental values for $(\Gamma_W$, $\Gamma_Z)$ as inputs.
However, since a number of SMEFT operators contribute to $\Gamma_V$, it is more appropriate to express $\Gamma_V$ using our other input parameters and expand it in powers of $\Lambda$. 
This is the approach adopted in~\cite{Brivio:2019myy}\footnote{We have verified that we have reasonable agreement with~\cite{Brivio:2019myy} at LO.} and gives a  narrow width approximation,
\begin{align}
\begin{split}
    \Gamma&_\mathrm{NWA}(\hsm \to 4f) = \Gamma(\hsm \to f\bar{f} V) \times \frac{\Gamma(V\to f\bar{f})}{\Gamma(V\to \mathrm{all})} \, \\
    &=\Gamma_\mathrm{SM, \mathrm{NWA}} + \frac{1}{\Lambda^2}\left[\Gamma_\mathrm{EFT}^{(0)} (f\bar{f}V) \times \mathrm{BR}^{(0)}_{\mathrm{SM}}(V\to f\bar{f}) + \Gamma_\mathrm{SM}^{(0)} (f\bar{f}V) \times \mathrm{BR}^{(0)}_{\mathrm{EFT}}(V\to f\bar{f})\right]\\
    &+\frac{1}{16\pi^2 \Lambda^2}\bigg[\Gamma_\mathrm{EFT}^{(1)} (f\bar{f}V) \times \mathrm{BR}^{(0)}_{\mathrm{SM}}(V\to f\bar{f}) + \Gamma_\mathrm{SM}^{(0)} (f\bar{f}V) \times \mathrm{BR}^{(1)}_{\mathrm{EFT}}(V\to f\bar{f})\\
    & \qquad \qquad  \quad \ + \Gamma_\mathrm{EFT}^{(0)} (f\bar{f}V) \times \mathrm{BR}^{(1)}_{\mathrm{SM}}(V\to f\bar{f}) + \Gamma_\mathrm{SM}^{(1)} (f\bar{f}V) \times \mathrm{BR}^{(0)}_{\mathrm{EFT}}(V\to f\bar{f})\bigg]\, ,
\end{split}
\end{align}
where the superscript is the loop order, and the branching ratios are expanded in loop orders and powers of $\Lambda$. 
Writing $(V \to f\bar{f})$ and $(V\to \mathrm{all})$ as $(f\bar{f})$ and $(\mathrm{all})$ for brevity, we have explicitly
\begin{align}
\begin{split}
    \mathrm{BR}_\mathrm{SM}(V \to f\bar{f}) =& \ \frac{\Gamma_\mathrm{SM}(f\bar{f})}{\Gamma_\mathrm{SM}(\mathrm{all})} \\
    =&\  \frac{\Gamma_\mathrm{SM}^{(0)}(f\bar{f})}{\Gamma_\mathrm{SM}^{(0)}(\mathrm{all})} +\frac{1}{16\pi^2}\left[\frac{\Gamma_\mathrm{SM}^{(1)}( f\bar{f})}{\Gamma_\mathrm{SM}^{(0)}(\mathrm{all})}-\frac{\Gamma_\mathrm{SM}^{(0)}(f\bar{f}) \times \Gamma_\mathrm{SM}^{(1)}(\mathrm{all})}{\left[\Gamma_\mathrm{SM}^{(0)}(\mathrm{all})\right]^2} \right]\\
    =& \ \mathrm{BR}^{(0)}_\mathrm{SM}(f\bar{f})+\frac{1}{16\pi^2}\mathrm{BR}^{(1)}_\mathrm{SM}(f\bar{f})\, ,
\end{split}
\end{align}
\begin{align}
\begin{split}
    \mathrm{BR}_\mathrm{EFT}(V &\to f\bar{f}) = \ \frac{1}{\Lambda^2}\mathrm{BR}^{(0)}_\mathrm{EFT}(f\bar{f})+\frac{1}{16\pi^2 \Lambda^2}\mathrm{BR}^{(1)}_\mathrm{EFT}(f\bar{f})\, \\
    =&\  \frac{1}{\Lambda^2}\left[\frac{\Gamma_\mathrm{EFT}^{(0)}(f\bar{f})}{\Gamma_\mathrm{SM}^{(0)}(\mathrm{all})} -\mathrm{BR}_\mathrm{SM}^{(0)}(f\bar{f}) \frac{\Gamma_\mathrm{EFT}^{(0)}(\mathrm{all})}{\Gamma_\mathrm{SM}^{(0)}(\mathrm{all})}\right]\\
    & +\frac{1}{16\pi^2\Lambda^2}\bigg[\frac{\Gamma_\mathrm{EFT}^{(1)}(f\bar{f})}{\Gamma_\mathrm{SM}^{(0)}(\mathrm{all})}-\mathrm{BR}_\mathrm{SM}^{(0)}(f\bar{f})\frac{\Gamma_\mathrm{EFT}^{(1)}(\mathrm{all})}{\Gamma_\mathrm{SM}^{(0)}(\mathrm{all})}-\mathrm{BR}_\mathrm{SM}^{(1)}(f\bar{f})\frac{\Gamma_\mathrm{EFT}^{(0)}(\mathrm{all})}{\Gamma_\mathrm{SM}^{(0)}(\mathrm{all})}\\
    & \ \qquad \qquad  \quad + 2\mathrm{BR}_\mathrm{SM}^{(0)}(f\bar{f}) \frac{\Gamma_\mathrm{EFT}^{(0)} (\mathrm{all})\Gamma_\mathrm{SM}^{(1)}(\mathrm{all}) }{\left[\Gamma_\mathrm{SM}^{(0)}(\mathrm{all})\right]^2} - \frac{\Gamma_\mathrm{EFT}^{(0)} (f\bar{f})\Gamma_\mathrm{SM}^{(1)}(\mathrm{all}) }{\left[\Gamma_\mathrm{SM}^{(0)}(\mathrm{all})\right]^2}\bigg] \, .
\end{split}
\end{align}

To implement this at LO without the NWA, we must write the widths $\Gamma_V$ in terms of $\Lambda$ and expand the total amplitude squared again:
\begin{align}
\begin{split}
    |\mathcal{A}(\hsm \to 4f)|^2 =& \ |\mathcal{A}_\mathrm{SM}|^2 + \frac{2}{\Lambda^2} \mathrm{Re} \left(\mathcal{A}_\mathrm{SM}^{(0)*} \times \mathcal{A}_\mathrm{EFT}^{(0)}  \right)\Bigg\vert_{\Gamma_V = \Gamma_\mathrm{SM}^{(0)}(V\to \mathrm{all})} \\ 
    & + \frac{\Gamma_\mathrm{EFT}^{(0)}(V\to \mathrm{all})}{\Lambda^2}\left(\frac{\partial}{\partial \Gamma_V}|\mathcal{A}_\mathrm{SM}|^2\right)\Bigg\vert_{\Gamma_V = \Gamma_\mathrm{SM}^{(0)}(V\to \mathrm{all})} \, .
\end{split}
\end{align}

It is important to estimate how well the NWA approximates the full results.
We can do this comparison only at LO, which should give a rough idea for what to expect at NLO.
For specific subprocesses, such as $\hsm \to 4\mu$, it can fail badly for specific coefficients, as shown in the next subsection. 
However, for most operators and individual $\hsm \to f_1 \bar{f}_2 f_3 \bar{f}_4$ subprocesses, the agreement is at the $\sim 10\%$ level.
In the total $\hsm \to 4f$ width, this remains true for bosonic operators and those entering $X_H$.
For fermionic operators, on the other hand, a delicate cancellation occurs between channels that results in the NWA and full result disagreeing at an $\mathcal{O}(1)$ level.
This can be seen from Table 9 of~\cite{Brivio:2019myy}, and is a manifestation of the fact that the sum of partial widths equals the total width, i.e. $\sum_X \Gamma(\hsm \to X) / \Gamma_V = 1$.
In the total $4f$ width, this cancellation is exact in the NWA, whereas in the full result off-shell and interference effects violate it.

\subsubsection{Differential distributions for \texorpdfstring{$\hsm \to 4\ell$}{H to 4l}}\label{sec:diff_dists}

As discussed in Sec.~\ref{sec:llZ}, we have implemented differential distributions in the dilepton invariant mass distributions,  $d\Gamma/dm_{\ell\ell}$, for the $\hsm \to 4\ell$ channel since they are especially relevant at the LHC.
These distributions are also particularly useful for analysing the validity of the NWA and how it fails.

To make a connection to the three-body $\hsm \rightarrow \ell^+\ell^-Z$ process, we define $m_Z^{\ell\ell}$ to be the opposite sign, same flavour dilepton invariant mass closest to $m_Z$, and $m_{Z^*}$ to be the opposite pair.
For example, for $\hsm \to e^+(p_1)e^-(p_2)\mu^+(p_3)\mu^-(p_4)$, $m^{\ell\ell}_Z = \mathrm{max}(m_{12},m_{34})$ and $m_{Z^*} = \mathrm{min}(m_{12},m_{34})$.
For the $4\mu$ and $4e$ channels, one must also consider $m_{14}$ and $m_{23}$.

\begin{figure}
	\centering  
    \hspace{-25pt}
        \includegraphics[width=0.49\textwidth]{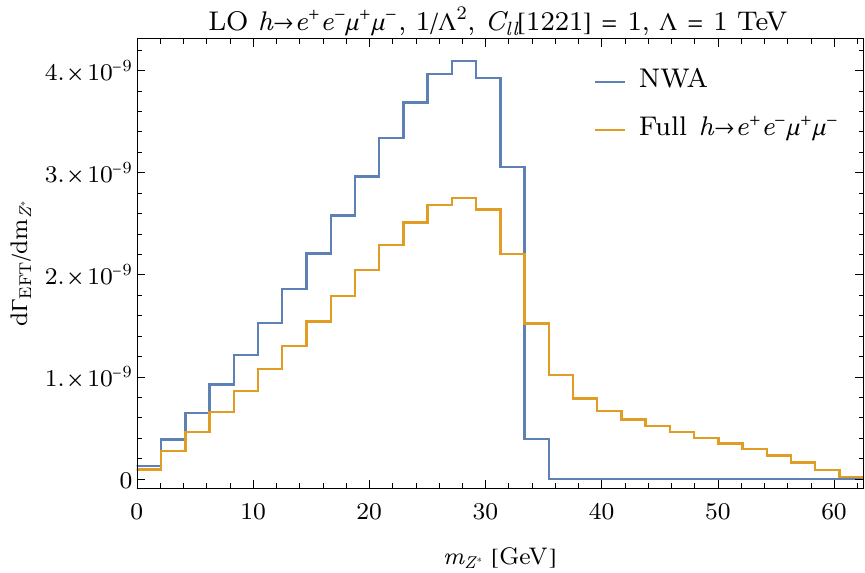}
        \includegraphics[width=0.49\textwidth]{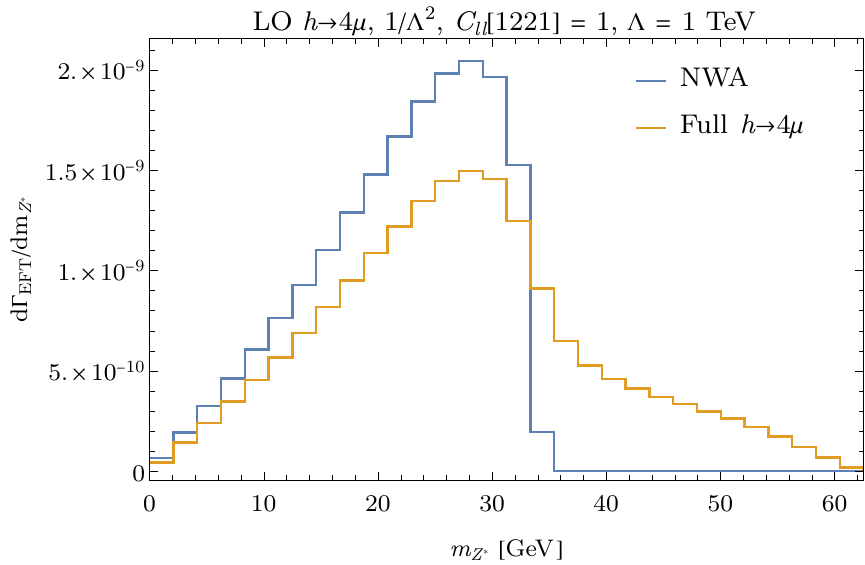}
	\caption{A comparison of the narrow width approximation and full $\hsm \to 4\ell$ results for $C_{ll}[1221]$, where the narrow width approximation works similarly for both $e^+e^-\mu^+\mu^-$ (left) and $4\mu$ (right) channels. Here $m_{Z^*}$ is defined to be the invariant mass of the opposite pair of same flavour opposite sign leptons from the pair with invariant mass closest to $M_Z$.} 
	\label{fig:4l_dists_Cll_1221}
\end{figure}

\begin{figure}
	\centering  
    \hspace{-25pt}
        \includegraphics[width=0.49\textwidth]{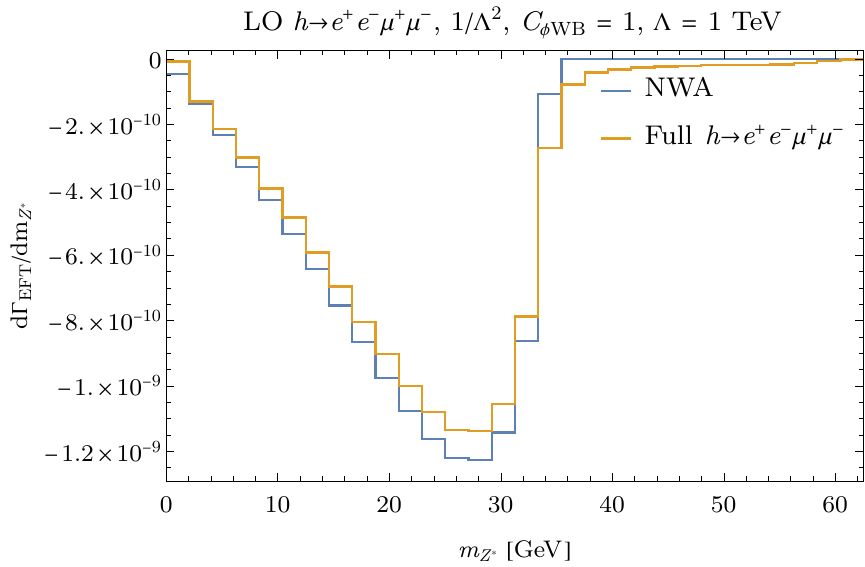}
        \includegraphics[width=0.49\textwidth]{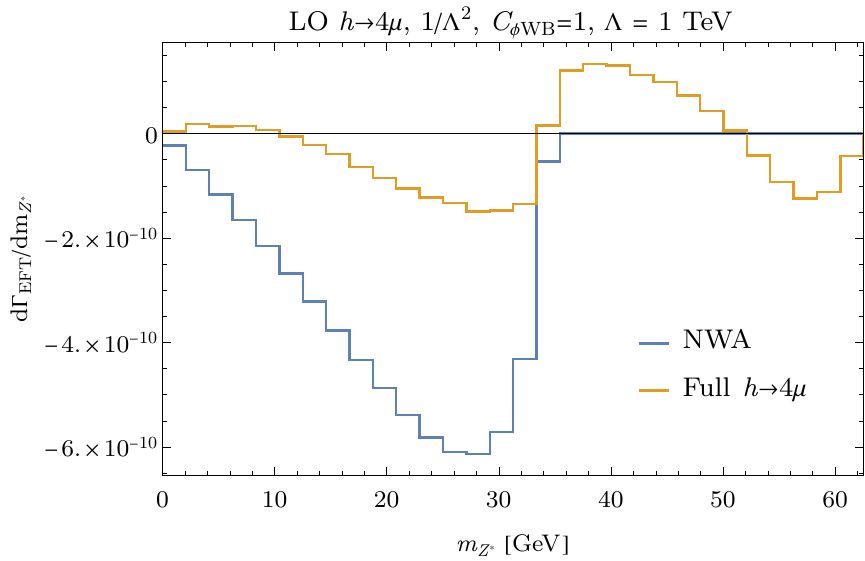}
	\caption{A comparison of the narrow width approximation and full $\hsm \to 4\ell$ results for $C_{\phi WB}$, where the narrow width approximation works for $e^+e^-\mu^+\mu^-$ (left) but fails badly for the $4\mu$ (right) channel. Here $m_{Z^*}$ is defined to be the invariant mass of the opposite pair of same flavour opposite sign leptons from the pair with invariant mass closest to $m_Z$.}
	\label{fig:4l_dists_CphiWB}
\end{figure}

We show results for two select coefficients in Figs.~\ref{fig:4l_dists_Cll_1221} and~\ref{fig:4l_dists_CphiWB} at LO for both the $e^+e^-\mu^+\mu^-$ and $4\mu$ modes.
As can be seen in Fig.~\ref{fig:4l_dists_Cll_1221}, the NWA works well inclusively for $C_{ll}[1221]$ in both channels, though at the level of the differential distribution the agreement is worse, particularly beyond $m_{Z^*} \gtrsim 34$ GeV.

For many operators in the $e^+e^-\mu^+\mu^-$ mode the agreement is actually quite good even differentially, as seen in Fig.~\ref{fig:4l_dists_CphiWB}.
On the other hand, for the $4\mu$ mode, neglected cross terms make the agreement worse, and it fails altogether for $C_{\phi WB}$ where virtual photon contributions become important. No other coefficients fail as dramatically as $C_{\phi WB}$, but the large differences should be kept in mind when using these results.

An important consequence of having access to $m_{\ell\ell}$ distributions is the ability to include the effect of the experimental cut $m_{Z^*} \ge 12$ GeV used in the ATLAS and CMS analyses~\cite{ATLAS:2023tnc,CMS:2025wnr,Dawson:2024pft}.
We include results with this cut in our numerical files at~\cite{GITLAB:newish}.

%% file: results.tex
In this section, we show some examples of the effects of NLO contributions to Higgs decays.  In general, the NLO contributions offer a window into the effects of operators that do not appear at tree level, extending the physics reach of the  HL-LHC program and future $e^+e^-$ colliders.
One of the major goals of these colliders is to measure the Higgs tri-linear coupling, as parameterized by $C_\phi$ in the SMEFT, which  first contributes to single Higgs production at one-loop order. It is hence important to understand the impact of a consistent calculation where all contributions are included at the same perturbative order. 

\subsection{HL-LHC projections}\label{sec:hl-lhc}

\begin{figure*}
	\centering  
    \hspace{-25pt}
        \includegraphics[width=0.44\textwidth]{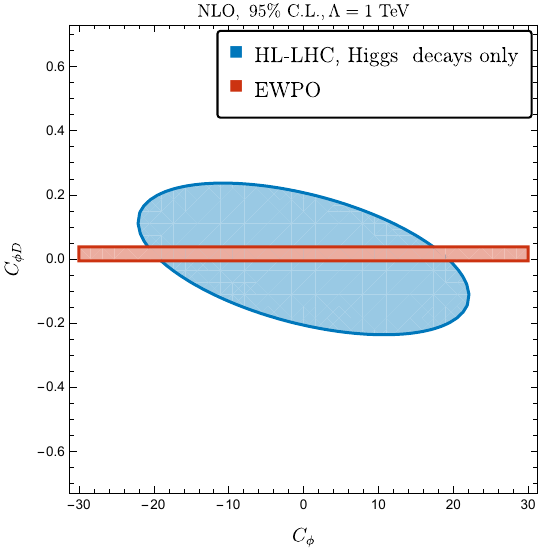}
         \hspace{40pt}
        \includegraphics[width=0.45\textwidth]{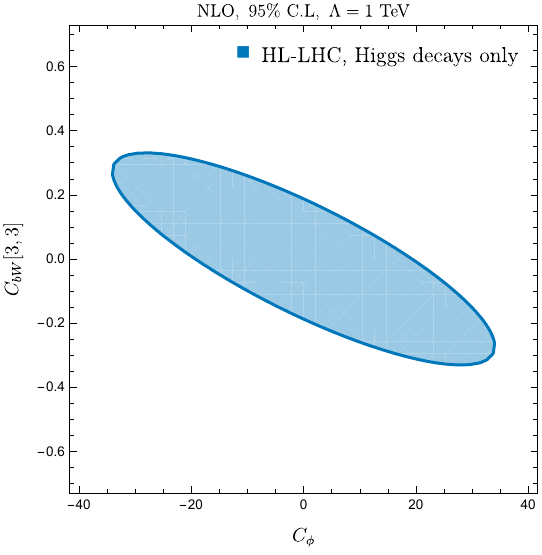}
	\caption{$95\%$ C.L. limits from a complete NLO electroweak calculation of EWPOs \cite{Bellafronte:2023amz}, compared with HL-LHC projected limits, including the Higgs decay at NLO electroweak. Note that the operators in the right-hand plot are not constrained by EWPOs.} 
	\label{fig:Cphilhc}
\end{figure*}

The inclusion of NLO SMEFT contributions provides sensitivity to operators that do not contribute to Higgs processes at LO.  Attention has focussed largely on $C_\phi$, which generates  anomalous Higgs tri-linear couplings \cite{Degrassi:2016wml,Degrassi:2017ucl,Gorbahn:2016uoy,Maura:2025rcv,terHoeve:2025omu}.  When the full suite of dimension-6 SMEFT operators are included at NLO, the results become highly correlated~\cite{DiVita:2017eyz}.  The complete set of calculations required to compute Higgs production and decay to NLO electroweak/QCD order in the dimension-6 SMEFT at the LHC does not yet exist and so we include Higgs production channels calculated in the SM, followed by the full NLO Higgs decay predictions in the dimension-6 SMEFT.  The results are hence only an indication of the impact of NLO electroweak corrections and motivate the complete calculation of production processes at NLO electroweak order for the LHC.

We assume that production and decay factorize at the HL-LHC (${\cal{L}}=3 ~\textrm{ab}^{-1}$)\footnote{This is not true for all dimension-6 operators.} and further set  the production processes to be given by the SM predictions.
The Higgs decays are included at NLO in the dimension-6 SMEFT as described in this paper and are parameterized as
\begin{equation}
\mu_f=\frac{BR (\hsm\rightarrow X_f)_{SMEFT}}{BR(\hsm\rightarrow X_f )_{SM}}\, ,
\end{equation}
where $X_f$ is the final state from the Higgs decay. We use the sensitivity projections of Ref.~\cite{ATL-PHYS-PUB-2018-054}, along with the updates of Ref.~\cite{atlas2025highlightshllhcphysicsprojections}. 
We show in Fig. \ref{fig:Cphilhc}  the regions where $\mu_{f}$, summed over all production and decay modes is within  $95\%$ C.L. of the SM. On the left-hand side of Fig. \ref{fig:Cphilhc}, we show the correlation of $C_\phi$ with $C_{\phi D}$ and on the right-hand side, the correlation of $C_\phi$ with $C_{bW} [33]$. The NLO accurate  results from EWPOs (using the data of Tab.~\ref{tab:expnums}) are also shown, and we see that the combination of HL-LHC with EWPOs  is highly restrictive. The right-hand side of Fig.  \ref{fig:Cphilhc} shows an example of two operators that do not contribute to EWPOs and demonstrates the strong correlation at HL-LHC.  We note, however, that the NLO EW/QCD SMEFT contributions to the production processes need to be included in future studies in order to have NLO consistent results.

\subsection{Comparing Tera-Z reach with Higgstrahlung Sensitivity}

The NLO SMEFT results for Higgstrahlung production, $e^+e^-\rightarrow Z\hsm$, can be combined with the NLO results for Higgs decays presented in this paper, and the NLO results for $Z$ decays from Refs.~\cite{Bellafronte:2023amz,Biekotter:2025nln}, to assess the expected sensitivities at future lepton colliders.  We assume a relative precision of $0.5\%$ for the total Higgstrahlung cross section at $\sqrt{s}=240~\textrm{GeV}$ and use the projected sensitivities for Higgs decays at FCC-ee from Tab. \ref{tab:fcc} to derive $95\% $ C.L. limits on pairs of dimension-6 SMEFT coefficients.  These limits can be compared  with those projected for the Tera-Z run, obtained from the electroweak precision observables listed in Tab.~\ref{tab:fcceenums}. To quantify the theoretical uncertainties we use the two scenarios defined in Ref.~\cite{deBlas:2944678}: a conservative one, in which the errors are projected according to what is likely to be achieved by extending the present computational tools, and an aggressive one, in which it is assumed that fundamental advances in theory techniques and tools will considerably reduce the related uncertainties for most of the observables. We report the detailed numbers in Tab.~\ref{tab:fcceenums}.

\begin{figure*}
	\centering  
    \hspace{-25pt}
        \includegraphics[width=0.44\textwidth]{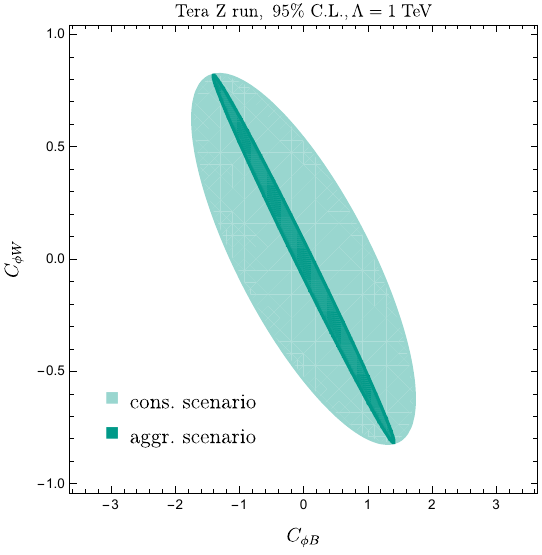}
         \hspace{40pt}
        \includegraphics[width=0.45\textwidth]{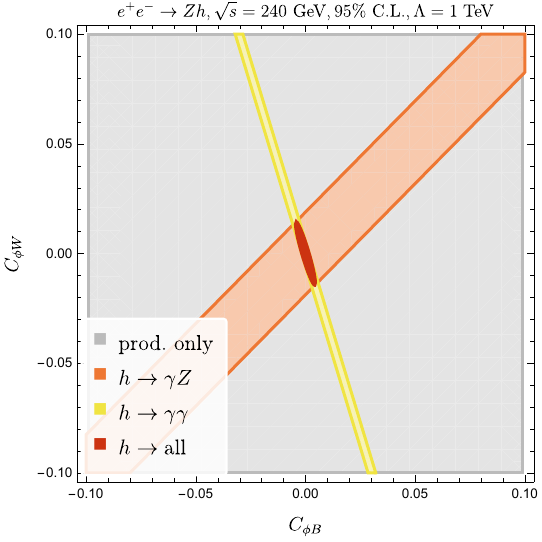}
	\caption{$95\%$ confidence level (C.L.) sensitivities to $C_{\phi B}$ and $C_{\phi W}$, for the Tera-Z run assumming either a conservative or aggressive set of theory uncertainties(left) and $Z\hsm$ production followed by the Higgs decays at $\sqrt{s}=240$ GeV (right).} 
    	\label{fig:CphiB-CphiW}
\end{figure*}

We begin by considering in Fig.~\ref{fig:CphiB-CphiW} pairs of operators that contribute to Higgstrahlung at tree-level, but enter $Z$ production at the peak only at one-loop, $C_{\phi B}$ and $C_{\phi W}$. 
The left-hand- side of the figure shows the limits derived at the Tera-Z run by adopting a conservative rather than an aggressive estimate of the theoretical uncertainties. 
We note the sensitivity of the result to the assumptions made about future theoretical uncertainties.
The combined information from Higgs production and decays shown on the right-hand side allows an improvement of at least one order of magnitude on these sensitivities, two in the case of $C_{\phi B}$. 
In particular, the right-hand plot shows the complementary constraints given from the decay channels $\hsm\to \gamma Z$ and $\hsm \to \gamma \gamma$, where the $O_{\phi B}$ operator enters at leading order, while the coefficient of $O_{\phi W}$, which first appears at loop level, takes a different sign in the $\gamma \gamma$ case, as compared to the $\gamma Z$ one.

\begin{figure*}
	\centering  
    \hspace{-25pt}
        \includegraphics[width=0.475\textwidth]{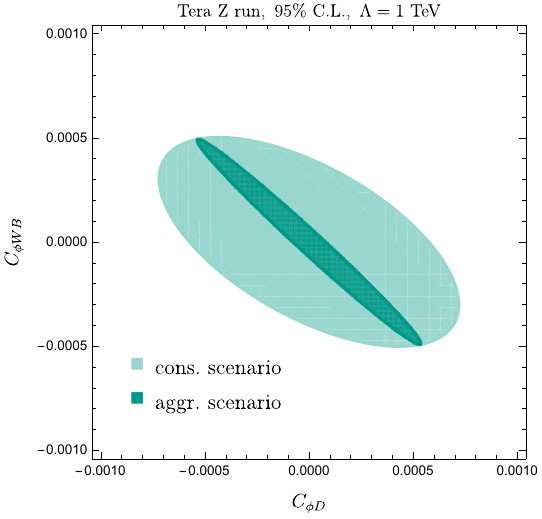}
         \hspace{40pt}
        \includegraphics[width=0.45\textwidth]{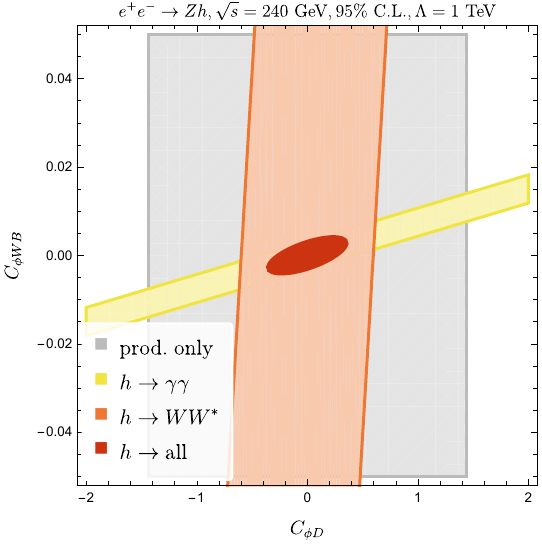}
	\caption{Same as Fig.~\ref{fig:CphiB-CphiW}, but with $95\%$ C.L. limits on $C_{\phi D}$ and $C_{\phi W B}$.} 
	\label{fig:CphiD-CphiWB}
\end{figure*}

Fig.~\ref{fig:CphiD-CphiWB} illustrates the sensitivity to two operators entering at leading order  in $Z$ production, $C_{\phi D}$ and $C_{\phi W B}$. In this case the higher statistics available at the Tera-Z run allows us to derive strong limits and it is apparent how tagging the Higgs decays enhances the constraining power of Higgstrahlung production alone. The $\hsm \to W W^*$ channel, computed here at NLO EW in the SMEFT for the first time, provides an order of magnitude improvement to the bound on $C_{\phi D}$ as compared to that found from production only. However, the Higgstrahlung process is not competitive with the Tera-Z results for these operators.
\begin{figure*}
	\centering  
    \hspace{-25pt}
        \includegraphics[width=0.45\textwidth]{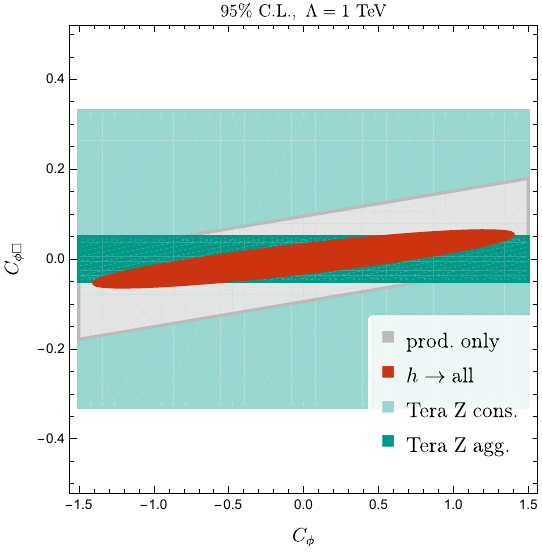}
         \hspace{40pt}
        \includegraphics[width=0.45\textwidth]{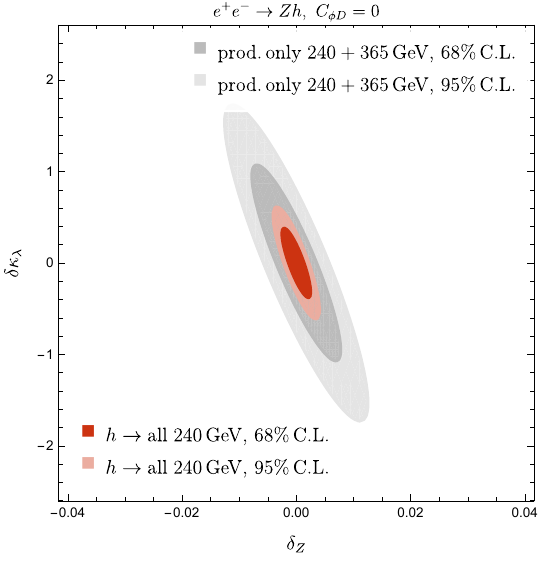}
	\caption{Bounds on coefficients contributing to $\hsm ZZ$/$\hsm WW$ and trilinear Higgs couplings. Left: SMEFT calculation in terms of $C_\phi$ and $C_{\phi \Box}$, with $95\%$ C.L. sensitivities from Higgsstrahlung with (without) the Higgs decays in red (grey) and from the Tera-Z run in green with aggressive and conservative assumptions about theory uncertainties. Right: Interpretation in terms of the $\delta_Z$ and $\delta \kappa_\lambda$ parameters computed at linear order, considering $C_{\phi D}=0$. Limits at $68\%$ and $95\%$ C.L. are given by combining by combining two energy runs at $\sqrt{s}=240$ and $365$ GeV (grey) with only the total cross section included and by considering Higgstrahlung with Higgs decays included at a single energy, $\sqrt{s}=240$ GeV(red/pink).}
	\label{fig:Cphi-CphiBox}
\end{figure*}
In Fig.~\ref{fig:Cphi-CphiBox} we present the sensitivity to $C_\phi$ and $C_{\phi \Box}$ on the left and to the parameters $\delta_Z$ and $\delta \kappa_\lambda$ on the right. 
The linear dimension-6 SMEFT calculation has been reinterpreted in terms of $\delta_Z$ and $\delta \kappa_\lambda$, parameterizing the deviations due to possible new physics in the $hVV$ couplings and the Higgs self-interaction~\cite{McCullough:2013rea,Maltoni:2018ttu,Altmann:2025feg,Maura:2025rcv,Rossia:2023hen,terHoeve:2025omu}, via the relations \cite{Asteriadis:2024xts}
\begin{eqnarray}
\delta_Z &=& \frac{1}{4} \frac{v^2}{\Lambda^2} \biggl(C_{\phi D}+4 C_{\phi\square}\biggr) \nonumber \\
\delta \kappa_\lambda &=& \frac{v^2}{\Lambda^2} \left( 3\biggl[\frac{C_{\phi D}}{4}- C_{\phi \square}\biggr] - 2  \frac{v^2}{\mhsm^2} C_\phi \right) \, ,
\end{eqnarray}
where we set $C_{\phi D} = 0$. On the left-hand side, the $95\%$ C.L. limits are derived by considering the Tera-Z run in green, and Higgsstrahlung production at $\sqrt{s}=240$~GeV in grey. Combining the latter with the information from Higgs decays is crucial for constraining $C_\phi$. On the right-hand side, we show the bounds obtained from production only by a joint fit of the runs at $\sqrt{s} = 240$ and $365$ GeV, where we assume
a relative precision of $1\%$ for the total Higgstrahlung cross section at $\sqrt{s} = 365$ GeV. Lighter colors correspond to $95\%$ C.L. and darker ones to $68\%$ C.L.
The red circles show how, when adding Higgs decay tagging on top of production, it is sufficient to run at a single energy  to obtain much more robust limits than from total cross section measurements alone.
This plot can be compared to Fig.~4 of Ref.~\cite{Bellafronte:2025jbk}, which is obtained by setting $C_{\phi \Box}=0$ and plotting $C_{\phi D}$ vs. $C_\phi$. Given that $C_{\phi \Box}$ is very well constrained, the derived limits on $\delta_Z$ and $\delta \kappa_\lambda$ are more stringent in the $C_{\phi \Box}-C_\phi$ fit.

For operators that contribute to $e^+e^-b{\overline{b}}$ and $e^+e^-t{\overline{t}}$ interactions at tree level,    the measurements of $e^+e^-\rightarrow b{\overline{b}}$ above the $Z$ pole and of $e^+e^-\rightarrow t{\overline{t}}$ at the $t {\overline{t}}$ threshold can significantly reduce the allowed parameter space for these coefficients~\cite{Greljo:2024ytg,Ge:2024pfn,Bellafronte:2025ubi,Allwicher:2025mvd}.  Interestingly, a measurement of $R_t$ at the $t {\overline{t}}$ threshold effectively eliminates the correlation between $C_\phi$ and the five $e^+e^-b{\overline{b}}$ and $e^+e^-t{\overline{t}}$ operators~\cite{Allwicher:2025mvd}.

\subsection{Case Study: Scalar Singlet Model}
It is interesting to examine the effects of NLO corrections to the Higgstrahlung process and compare them with those resulting from  Tera-Z data.  As a test case,
we study the BSM model with a single heavy neutral scalar particle, $S$, added to the SM~\cite{Dawson:2020oco,Gorbahn:2015gxa,Robens:2016xkb}.    
This model serves to illustrate many of the assumptions that go into deriving conclusions about the discovery reach of future colliders. The scalar singlet model is highly motivated since it can generate a first order electroweak phase transition for some choices of the parameters~\cite{Chen:2017qcz,Carena:2019une}.

The most general potential coupling a gauge singlet scalar, $S$, to the SM Higgs doublet, $\Phi^T=(\phi^+, \phi^0)$ is,
   \begin{align}
V(\Phi, S) = & -\mu_H^2\Phi^\dagger\Phi+\lambda_H(\Phi^\dagger\Phi)^2  + \frac{m_\xi}{2}\Phi^\dagger\Phi S+\frac{\kappa}{2}\Phi^\dagger\Phi S^2 \nonumber \\
&  + t_S S+\frac{M^2}{2}S^2+\frac{m_\zeta}{3}S^3 +\frac{\lambda_S}{4}S^4\, .
\end{align} 
There are two distinct cases:  The model possesses a $\mathbb{Z}_2$ symmetry, in which case $m_\zeta=t_S=m_\xi=0$ or there is no $\mathbb{Z}_2$ symmetry in which case $S$ can obtain a non-zero vev and the Lagrangian parameters can be shifted to eliminate the tadpole term, $t_S$.  The physical scalars, $h_1$ and $h_2$, are admixtures of $S$ and $\phi^0$, and we will assume that $h_1=\hsm$ is the SM Higgs boson, while $h_2$ is much heavier.

It is straightforward to match the singlet model to the dimension-6 SMEFT  at tree- level by taking the limit $m_{h_2}\rightarrow \infty$.  In this limit, $m_{h_2}\rightarrow M$ and two SMEFT coefficients are generated at tree level,
\begin{align}
O_{\phi}=&|\Phi^\dagger\Phi|^3\\
O_{\phi \square}=&|\Phi ^\dagger\Phi|\square |\Phi ^\dagger\Phi|\, .
\end{align}

\begin{figure}[t]
 \centering
 \includegraphics[width=0.49\textwidth]{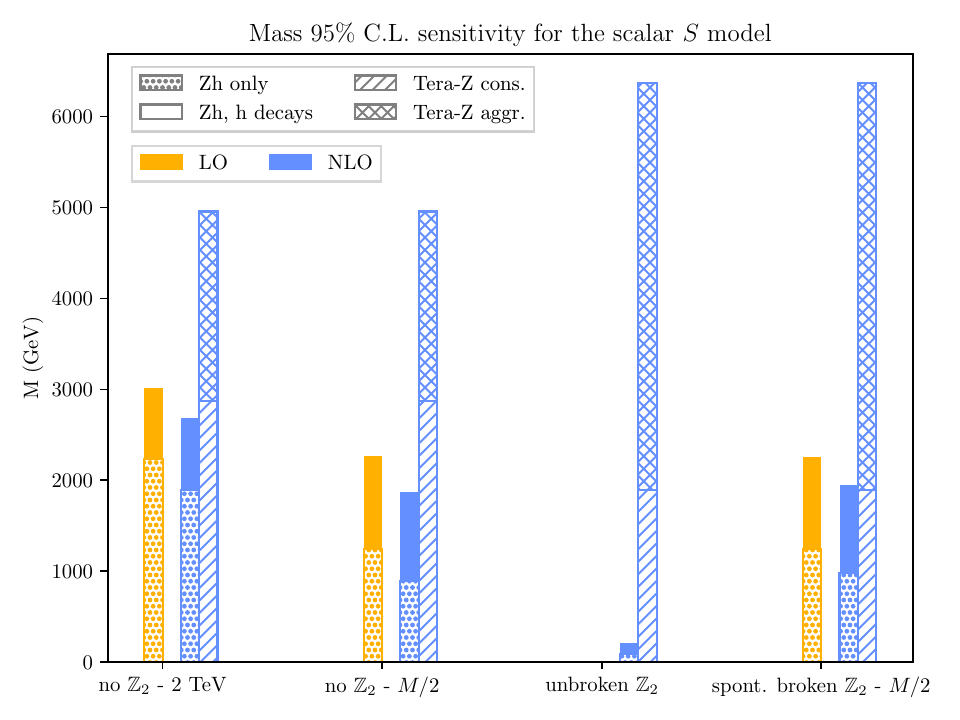}
 \includegraphics[width=0.49\textwidth]{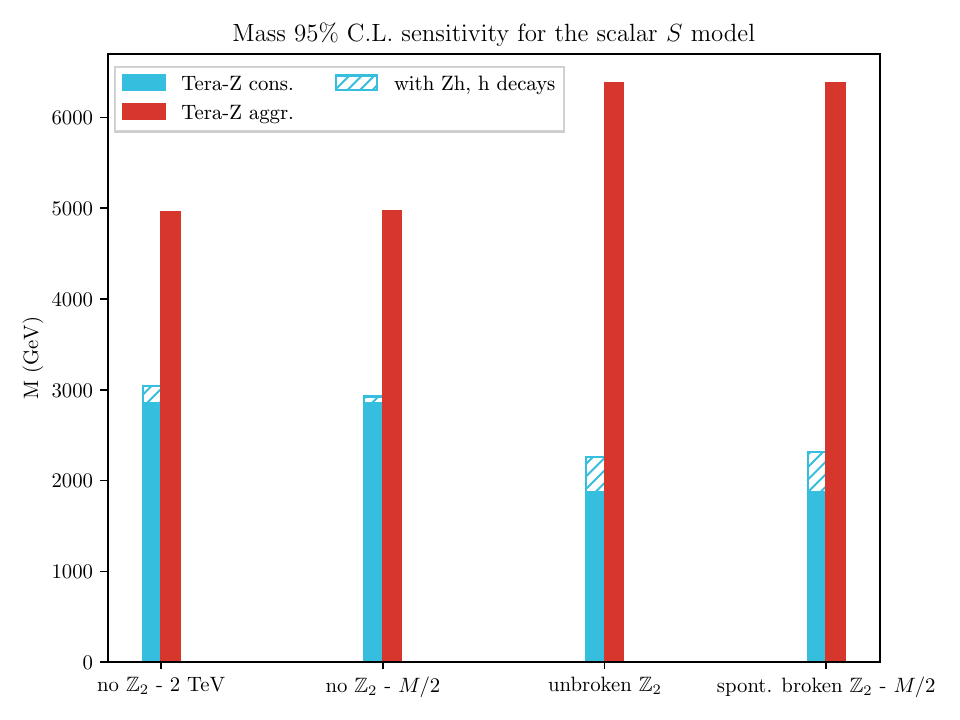}
 \caption{$95\%$ C.L. mass reach for the $S$ model under the different symmetry hypotheses discussed in the text. The labels $2$ TeV and $M/2$ refer to the assumptions on dimensional parameters discussed in the main text. Left: $e^+ e^- \to Z\hsm$ at $\sqrt{s}=240$ GeV (dots), Higgsstrahlung combined with Higgs decays (solid), the Tera-Z data in the conservative (stripes) and aggressive scenarios (crosses). The matching is performed at LO (yellow) and NLO (blue).
 Right: NLO matching considering the Tera-Z conservative (light blue) and aggressive (red) uncertainty scenarios combined with the run at $\sqrt{s}=240$ GeV including Higgsstrahlung and the Higgs decays (stripes). \label{fig:massreach}}
 \end{figure}

There are 3 possibilities for the pattern of coefficients~\cite{Gorbahn:2015gxa,Jiang:2018pbd,Haisch:2020ahr}:
\begin{itemize}
\item  The $\mathbb{Z}_2$ symmetry remains unbroken in which case no operators are generated at tree- level.  At one-loop level, $O_\phi$ and $O_{\phi \square}$ are generated with coefficients:
\begin{align}
    C_{\phi}=&-\frac{\kappa^3}{192 \pi^2 } \nonumber \\
    C_{\phi {\square}}=&-\frac{\kappa^2}{384\pi^2 }\, .
\end{align}
\item The  $\mathbb{Z}_2$ symmetry can be spontaneously broken and the $\mathbb{Z}_2$ symmetry implies the relationship, $\kappa=\frac{m_\zeta m_\xi}{3 M^2}$. At tree- level: 
\begin{align}
    C_{\phi}=& 0\nonumber \\
    C_{\phi {\square}}=&-\frac{m_\xi^2}{8M^2}\, .
\end{align}
\item If there is no $\mathbb{Z}_2$ symmetry, the tree level contributions are,
\begin{align}
    C_{\phi}=&\frac{m_\xi^2}{8M^2} \biggl[\frac{m_\zeta m_\xi}{3M^2}-\kappa\biggr]\nonumber \\
    C_{\phi {\square}}=&-\frac{m_\xi^2}{8M^2}\, .
\end{align}
\end{itemize}

In the numerical results for the matching, we set all dimensionless parameters in the UV-complete model to one, while discussing possible choices for the dimensional quantities.

Fig.~\ref{fig:massreach} shows the $95\%$ C.L. mass reach for the $S$ model under these different symmetry assumptions. In the left panel, dotted columns refer to fitting only the total cross section for $e^+ e^- \to Z\hsm$ at $\sqrt{s}=240$ GeV, while solid colors correspond to Higgstrahlung production combined with Higgs decays. Single- and double-striped hatches denote the Tera-Z run in the conservative and aggressive uncertainty scenarios, respectively. The colors identify the order of the matching: LO in yellow and NLO in blue. 
For consistency, the accuracy of the observables used in this analysis is always the same as the matching procedure.

Let us first consider the case with the no $\mathbb{Z}_2$ symmetry hypothesis. To translate the matching equations into mass limits, we always set dimensionless parameters of the UV model to one, and then need to make some assumption on the dimensional coefficients of the complete theory. In the literature, common choices are $m_\xi/2 = m_\zeta/3 = 1$ TeV or $m_\xi/2 = m_\zeta/3 =2$ TeV, and we present the latter in the plot. The resulting mass reach is however not always higher than $2$ TeV, invalidating the expansion in $m/M$, with $m$ a generic dimensional coefficient of the complete model, that is assumed in the matching. We therefore adopt the assumption $m_\xi/2 = m_\zeta/3 = M/2$ and fit $M$. The limits derived within this hypothesis are considerably different from the ones obtained by setting $m_\xi/2 = m_\zeta/3 =2$ TeV, if the matching is not performed considering the NLO complete calculation.  This demonstrates the impact of assumptions about the dimensional couplings on the projected mass reach limits.

The case of spontaneously broken $\mathbb{Z}_2$ symmetry is similar, because the fit is driven by $C_{\phi \Box}$, which is much more constrained than $C_\phi$, as can be seen in the left-hand panel of Fig.~\ref{fig:Cphi-CphiBox}. In addition, Tera-Z data has no sensitivity to $C_\phi$.
In general, one can observe that the information from Higgs production combined with the decays and from the Tera-Z data provide similar bounds, since the higher statistics available at the $Z$ peak compensates the lack of sensitivity on $C_\phi$.

Finally,  we assume that the $\mathbb{Z}_2$ symmetry is left unbroken, and $O_\phi$ and $O_{\phi \Box}$ are generated first at one- loop. In this case, the higher statistics from the $Z$ peak gives the strongest limits on the $S$ model.

The right-hand panel of Fig.~\ref{fig:massreach} shows the mass reach once the data from the  Tera-Z run have been combined with the data coming from Higgsstralung plus decays. As expected from their relative size, the combination of the bounds from Higgs processes with those of the conservative scenario yields a slight improvement with respect to that of the Tera-Z run alone. On the other hand, no appreciable improvement from the Higgstrahlung data can be seen in the aggressive theory scenario.

%% file: conc.tex
We have computed all two- and three-body Higgs decays at NLO in the dimension-6 SMEFT.  The physical
four-body final states $\hsm\rightarrow (f_1{\bar{f}}_2)(f_3 {\bar{f_4}})$ are determined by using the exact four-body SMEFT result for the LO contribution and including the NLO contribution by combining the NLO $\hsm\rightarrow V (f_1 {\bar{f_2}})$ result with the NLO result for $V\rightarrow 
(f_3 \bar{f_4})$  using the NWA. The results are contained in the publicly available Monte Carlo code {\sc NEWiSH}~\cite{GITLAB:newish} and numerical results are provided both in the text and at the \href{https://gitlab.com/mforslund/newish}{GitLab~\faGitlab}~repository, allowing for the NLO EW/QCD evaluation of Higgs branching ratios to ${\cal{O}}\left(\frac{1}{(4 \pi \Lambda)^2}\right)$ in the dimension-6 SMEFT. The most up-to-date results will always be found in the GitLab repository.

The effects of NLO corrections are largely to induce correlations between operator contributions to observables, implying that great care is required in interpreting single parameter limits.  We demonstrate that by combining results from Tera-Z with data from Higgstrahlung, many of these ambiguities can be resolved. For operators that contribute to Tera-Z at tree level, including the Higgstrahlung data with Higgs decays to NLO EW order gives only a small improvement to the sensitivity from Tera-Z results alone in the cases we have studied.  However, for operators that do not contribute to Tera-Z at tree level, the Higgstrahlung measurements combined with the NLO Higgs decays  can result in   significantly improved sensitivity. We can therefore anticipate interesting effects when incorporating the results presented here into global analyses.

Most of our conclusions depend sensitively on the assumptions about SM theory uncertainties, demonstrating the need for more precise theoretical calculations. In a similar fashion as in Ref.~\cite{deBlas:2944678}, we considered two scenarios for the theoretical uncertainties of the future Tera-Z run, corresponding to different levels of theoretical progress and technological advancement. As a test case, we examined the Tera-Z reach for a scalar singlet and observed that the mass bound differs depending  on the assumptions about the $Z_2$ symmetry of the potential and the numerical values of the dimensional parameters. In this example, the Higgstrahlung data slightly improve the projected mass reach obtained from the Tera-Z run only in the conservative scenario for theory uncertainties, while no meaningful improvement from Higgstrahlung measurements can be observed in the aggressive scenario.

These results highlight an important point. Under conservative assumptions, the Tera-Z run offers only limited improvement over other FCC-ee runs in probing BSM physics. By contrast, the aggressive scenario is expected to deliver markedly greater advances. This clearly demonstrates the need for the theoretical community to push towards the development of new technologies and methods in order to reach the level of precision required to fully take advantage of the FCC-ee. 

The extraction of SMEFT coefficients from HL-LHC data is a crucial tool in the search for high scale new physics.  Our work represents an important step toward NLO electroweak/QCD accurate predictions in the dimension-6 SMEFT, but a complete calculation including Higgs production processes at the LHC is needed for accurate global fits.  Furthermore, the complete NLO calculation of $h\rightarrow 4f$ without using the NWA is also a necessary ingredient for increasing the precision of future SMEFT studies. 

 Note added:  The numerical results of this work have been presented in the POPxf format \cite{bellafronte2026nloewqcddimension6}.

%% file: code.tex
To make our results more easily accessible and replicable, we have written a publicly available code {\sc NEWiSH}~\href{https://gitlab.com/mforslund/newish}{\faGitlab}~\cite{GITLAB:newish} (\textbf{N}LO \textbf{E}lectro\textbf{W}eak \textbf{i}n the \textbf{S}MEFT for \textbf{H}iggs widths).
{\sc NEWiSH} is a fixed-order Fortran code that computes the partial widths for the two- and three- body decay channels presented in this work at NLO accuracy in the SMEFT consistently retaining contributions to ${\cal{O}}(\frac{1}{16\pi^2\Lambda^2})$.
{\sc NEWiSH} also implements the full $\hsm \to 4f$ processes at LO, $\mathcal{O}\left(\frac{1}{\Lambda^2}\right)$.

{\sc NEWiSH} relies on the following dependencies:
\begin{itemize}
    \item \href{https://github.com/jacobwilliams/json-fortran}{json-fortran}
    \item \href{https://collier.hepforge.org/}{{\sc Collier}}~\cite{Denner:2016kdg} for evaluating loop integrals
    \item \href{https://feynarts.de/cuba/}{{\sc CUBA}}~\cite{Hahn:2004fe} for numerical integration
\end{itemize}

All processes have been validated using {\sc MadGraph}~\cite{Alwall:2014hca} at NLO QCD and EW in the SM, as well at LO in the SMEFT for a number of coefficients using SMEFTsim~\cite{Brivio:2020onw}.

We use the WCxf format~\cite{Aebischer:2017ugx} for Wilson coefficient inputs, though we should emphasize that while our coefficients are evaluated at the scale $\mu=\mhsm$ the code does \textbf{not} perform the running between the $\Lambda$ and $\mu$ scales needed for matching to specific UV models. 
This must be done independently using, for example, {\sc DsixTools}~\cite{Celis:2017hod,Fuentes-Martin:2020zaz}.

For up-to-date details on installation and usage of {\sc NEWiSH}, we refer to the READMEs on the GitLab page. 






%% file: hff.tex
\subsection{\texorpdfstring{$\hsm\rightarrow l^+l^-$}{H to ll}}\label{sewc:apphff}
 
Using the inputs of Sec.~\ref{sec:inputs}, 
\begin{align}
\begin{split}
    \Gamma_\textrm{LO}&(\hsm\rightarrow \tau^+\tau^-) = 0.259 \times 10^{-3} \ \textrm{GeV} \\
    &+\left(\frac{1 \ \textrm{TeV}}{\Lambda}\right)^2 \bigg(0.0314 C_{\phi\square}-0.00785 C_{\phi D}-3.08 C_{e\phi}[33]+0.0157 C_{ll}[1221]\\
    &-0.0157 C_{\phi l}^{(3)}[11]-0.0157 C_{\phi l}^{(3)}[22]\bigg) \times 10^{-3} \ \textrm{GeV}
\end{split}
\end{align}
\begin{align}
\begin{split}
    \delta\Gamma&_\textrm{NLO}(\hsm\rightarrow \tau^+\tau^-) =-2.58 \times 10^{-6} \ \textrm{GeV} \\
    &+\left(\frac{1 \ \textrm{TeV}}{\Lambda}\right)^2 \bigg(0.374 C_{\phi} +2.41 C_{\phi B}+0.567 C_{\phi\square}+0.599 C_{\phi D}+1.06 C_{\phi W}\\
    &+0.213 C_{\phi WB}+0.017 C_{d\phi }[33]+2.62 C_{eB}[33]-67 C_{e\phi }[33]-9.58 C_{eW}[33]\\
    &-0.154 C_{le}[3333]+0.857 C_{ledq}[3333]-0.262 C_{lequ}^{(1)}[3322]-54.9 C_{lequ}^{(1)}[3333]\\
    &-0.345 C_{ll}[1122]-0.201 C_{ll}[1221]+0.0428 C_{lq}^{(3)}[1133]+0.0428 C_{lq}^{(3)}[2233]\\
    &-0.0584 C_{\phi e}[33]-0.016 C_{\phi l}^{(1)}[11]-0.016 C_{\phi l}^{(1)}[22]+0.067 C_{\phi l}^{(1)}[33]\\
    &-0.0598 C_{\phi l}^{(3)}[11]-0.0598 C_{\phi l}^{(3)}[22]+0.28 C_{\phi l}^{(3)}[33]-0.0856 C_{\phi q}^{(3)}[33]\\
    &-0.00583 C_{\phi ud}[33]-0.711 C_{u\phi}[33]\bigg) \times 10^{-6} \ \textrm{GeV}
\end{split}
\end{align}
where we have dropped terms contributing less than $10^{-9}$ GeV.
Note the change from $10^{-3}$ to $10^{-6}$ GeV in the NLO correction for ease of presentation.

For $\hsm\rightarrow \mu^+\mu^-$, 
\begin{align}
\begin{split}
    \Gamma_\textrm{LO}&(\hsm\rightarrow \mu^+\mu^-) = 0.917\times 10^{-6} \ \textrm{GeV} \\
    &+\left(\frac{1 \ \textrm{TeV}}{\Lambda}\right)^2 \bigg(0.111 C_{\phi \square}-0.0278 C_{\phi D}-183  C_{e\phi}[22]\\
    &+0.0556 C_{ll}[1221]-0.0556 C_{\phi l}^{(3)}[11]-0.0556 C_{\phi l}^{(3)}[22]\bigg) \times 10^{-6} \ \textrm{GeV}
\end{split}
\end{align}
\begin{align}
\begin{split}
    \delta\Gamma_\textrm{NLO}&(\hsm\rightarrow \mu^+\mu^-) = -0.0278 \times 10^{-6} \ \textrm{GeV} \\
    &+\left(\frac{1 \ \textrm{TeV}}{\Lambda}\right)^2 \bigg(0.00133 C_{\phi }+0.0204 C_{\phi B}+0.00467 C_{\phi D}+0.00721 C_{\phi W}\\
    &-0.00179 C_{\phi WB}+0.157 C_{eB}[22]-2.11 C_{e\phi }[22]-0.571 C_{eW}[22]\\
    &-0.00919 C_{le}[2332]+0.00103 C_{ledq}[2222]+0.051 C_{ledq}[2233]\\
    &-0.0156 C_{lequ}^{(1)}[2222]-3.27 C_{lequ}^{(1)}[2233]-0.00122 C_{ll}[1122]\\
    &-0.00298 C_{ll}[1221]+0.00206 C_{\phi l}^{(3)}[11]+0.00305 C_{\phi l}^{(3)}[22]\\
    &-0.00252 C_{u\phi }[33]\bigg) \times 10^{-6} \ \textrm{GeV}
\end{split}
\end{align}
where we have dropped contributions smaller than $10^{-9}$ GeV.

\subsection{\texorpdfstring{$\hsm\rightarrow q {\overline{q}}$}{H to qq}}

\subsubsection{On-shell inputs}

With the inputs of Sec.~\ref{sec:inputs} and using OS renormalization for the fermion masses, 
\begin{align}
\begin{split}
    \Gamma_\textrm{LO}&(\hsm\rightarrow b\bar{b}) = 5.91 \times 10^{-3} \ \textrm{GeV}\\
    &+\left(\frac{1 \ \textrm{TeV}}{\Lambda}\right)^2\bigg(0.716 C_{\phi \square}-0.179 C_{\phi D}-25.3 C_{d\phi}[33]+0.358 C_{ll}[1221]\\
    &-0.358 C_{\phi l}^{(3)}[11]-0.358 C_{\phi l}^{(3)}[22]\bigg) \times 10^{-3} \ \textrm{GeV}
\end{split}
\end{align}
\begin{align}
\begin{split}
    \delta\Gamma&_\textrm{NLO}(\hsm\rightarrow b\bar{b}) = -2.25 \times 10^{-3} \ \textrm{GeV} \\
    &+\left(\frac{1 \ \textrm{TeV}}{\Lambda}\right)^2 \bigg(0.00847 C_{\phi} +0.00274 C_{\phi B}-0.253 C_{\phi \square}+0.0653 C_{\phi D}\\
    &+0.922 C_{\phi G}+0.0097 C_{\phi W}+0.00105 C_{\phi WB}+0.00211 C_{dB}[33]-0.132 C_{dG}[33]\\
    &+1.47 C_{d\phi }[33]+1.01 C_{dW}[33]+0.0000463 C_{e\phi }[33]+0.000847 C_{ledq}[3333]\\
    &-0.00788 C_{ll}[1122]-0.138 C_{ll}[1221]-1.89\times 10^{-6} C_{ll}[1331]\\
    &-1.89\times 10^{-6} C_{ll}[2332]
    +0.000976 C_{lq}^{(3)}[1133]+0.000976 C_{lq}^{(3)}[2233]\\
    &-0.000717 C_{\phi d}[33]-0.000366 C_{\phi l}^{(1)}[11]-0.000366 C_{\phi l}^{(1)}[22]+0.132 C_{\phi l}^{(3)}[11]\\
    &+0.132 C_{\phi l}^{(3)}[22]+4.26\times 10^{-6} C_{\phi l}^{(3)}[33]+0.00212 C_{\phi q}^{(1)}[33]-0.0109 C_{\phi q}^{(3)}[33]\\
    &-0.0631 C_{\phi ud}[33]-0.00354 C_{qd}^{(1)}[3333]-0.00473 C_{qd}^{(8)}[3333]+0.528 C_{quqd}^{(1)}[3333]\\
    &+0.101 C_{quqd}^{(8)}[3333]-0.0108 C_{u\phi }[33]+0.0164 C_{uW}[33]\bigg) \times 10^{-3} \ \textrm{GeV}
\end{split}
\end{align}
where we have not dropped any contributions.

For $\hsm\rightarrow c\bar{c}$, 
\begin{align}
\begin{split}
    \Gamma_\textrm{LO}&(\hsm\rightarrow c\bar{c}) = 0.561\times 10^{-3} \ \textrm{GeV}\\
    &+\left(\frac{1 \ \textrm{TeV}}{\Lambda}\right)^2\bigg(0.068 C_{\phi\square}-0.017 C_{\phi D}+0.034 C_{ll}[1221]-0.034 C_{\phi l}^{(3)}[11]\\
    &-0.034 C_{\phi l}^{(3)}[22]-7.84 C_{u\phi}[22] \bigg) \times 10^{-3} \ \textrm{GeV}
\end{split}
\end{align}
\begin{align}
\begin{split}
    \delta\Gamma&_\textrm{NLO}(\hsm\rightarrow c\bar{c}) = -0.309\times 10^{-3} \ \textrm{GeV} \\
    &+\left(\frac{1 \ \textrm{TeV}}{\Lambda}\right)^2 \bigg(0.00081 C_{\phi }+0.00234 C_{\phi B}-0.0356 C_{\phi \square}+0.00966 C_{\phi D}\\
    &+0.15 C_{\phi G}+0.0016 C_{\phi W}+0.000153 C_{\phi WB}+0.0000369 C_{d\phi }[33]\\
    &+4.4\times 10^{-6} C_{e\phi }[33]-0.000262 C_{lequ}^{(1)}[3322]-0.000749 C_{ll}[1122]\\
    &-0.0184 C_{ll}[1221]+0.0000927 C_{lq}^{(3)}[1133]+0.0000927 C_{lq}^{(3)}[2233]\\
    &-0.0000347 C_{\phi l}^{(1)}[11]-0.0000347 C_{\phi l}^{(1)}[22]+0.0178 C_{\phi l}^{(3)}[11]+0.0178 C_{\phi l}^{(3)}[22]\\
    &-0.000174 C_{\phi q}^{(1)}[22]+0.000634 C_{\phi q}^{(3)}[22]-0.000185 C_{\phi q}^{(3)}[33]+0.0000983 C_{\phi u}[22]\\
    &-0.0000126 C_{\phi ud}[33]-0.000334 C_{qu}^{(1)}[2222]-0.178 C_{qu}^{(1)}[2332]\\
    &-0.000446 C_{qu}^{(8)}[2222]-0.237 C_{qu}^{(8)}[2332]+0.00219 C_{quqd}^{(1)}[2233]\\
    &+0.000364 C_{quqd}^{(1)}[3223]+0.000486 C_{quqd}^{(8)}[3223]-0.0037 C_{uB}[22]\\
    &-0.0416 C_{uG}[22]+1.13 C_{u\phi }[22]-0.00154 C_{u\phi }[33]\\
    &-0.0245 C_{uW}[22]\bigg) \times 10^{-3} \ \textrm{GeV}
\end{split}
\end{align}
where we have dropped terms contributing less than $10^{-9}$ GeV.

For $\hsm\rightarrow s\bar{s}$ 
\begin{align}
\begin{split}
    \Gamma_\textrm{LO}&(\hsm\rightarrow s\bar{s}) = 2.46\times 10^{-6} \ \textrm{GeV} \\
    &+\left(\frac{1 \ \textrm{TeV}}{\Lambda}\right)^2 \bigg(0.299 C_{\phi \square }-0.0747 C_{\phi D}-520 C_{d\phi}[22]+0.149 C_{ll}[1221]\\
    &-0.149 C_{\phi l}^{(3)}[11]-0.149 C_{\phi l}^{(3)}[22]\bigg) \times 10^{-6} \ \textrm{GeV}
\end{split}
\end{align}
\begin{align}
\begin{split}
    \delta\Gamma&_\textrm{NLO}(\hsm\rightarrow s\bar{s}) = -2.35\times 10^{-6} \ \textrm{GeV} \\
    &+\left(\frac{1 \ \textrm{TeV}}{\Lambda}\right)^2 \bigg(0.00356 C_{\phi} +0.00533 C_{\phi B}-0.277 C_{\phi \square}+0.0709 C_{\phi D}\\
    &+1.47 C_{\phi G}+0.00581 C_{\phi W}+0.0474 C_{dB}[22]-2.77 C_{dG}[22]+181. C_{d\phi }[22]\\
    &-1.62 C_{dW}[22]+0.00103 C_{ledq}[2222]+0.0174 C_{ledq}[3322]-0.00329 C_{ll}[1122]\\
    &-0.14 C_{ll}[1221]+0.138 C_{\phi l}^{(3)}[11]+0.138 C_{\phi l}^{(3)}[22]+0.00289 C_{\phi q}^{(3)}[22]\\
    &-0.00329 C_{\phi ud}[22]-0.00146 C_{qd}^{(1)}[2222]-0.0727 C_{qd}^{(1)}[2332]-0.00195 C_{qd}^{(8)}[2222]\\
    &-0.097 C_{qd}^{(8)}[2332]+0.0517 C_{quqd}^{(1)}[2222]+1.55 C_{quqd}^{(1)}[2332]+9.28 C_{quqd}^{(1)}[3322]\\
    &+0.00984 C_{quqd}^{(8)}[2222]+2.06 C_{quqd}^{(8)}[2332]-0.00676 C_{u\phi }[33]\bigg) \times 10^{-6} \ \textrm{GeV}
\end{split}
\end{align}
where we have dropped terms smaller than $10^{-9}$ GeV.

\subsubsection{\texorpdfstring{$\overline{\mathrm{MS}}$ inputs}{MSbar inputs}}

As discussed in Sec.~\ref{sec:msbar}, the choice of $\overline{\mathrm{MS}}$ inputs for the quark masses makes a significant difference for the $\hsm \to q\bar{q}$ processes.
Here we give the expressions for the $\overline{\mathrm{MS}}$ inputs in Sec.~\ref{sec:inputs} while keeping all other inputs the same.
This contribution is also implemented in {\sc NEWiSH} for these fixed inputs.
Note that for each of $b\bar{b}$, $c\bar{c}$, and $s\bar{s}$ there are numerous small contributions introduced in this scheme from the running of the quark mass from $\mu=m_q$ to $\mu=\mhsm$ which we neglect here for brevity.

For $\hsm \to b\bar{b}$ we have
\begin{align}
\begin{split}
    \Gamma^{\overline{\mathrm{MS}}}_\textrm{LO}&(\hsm\rightarrow b\bar{b}) = 6.58 \times 10^{-3} \ \textrm{GeV}\\
    &+\left(\frac{1 \ \textrm{TeV}}{\Lambda}\right)^2\bigg(0.798 C_{\phi \square}-0.199 C_{\phi D}-26.7 C_{d\phi }[33]+0.399 C_{ll}[1221]\\
    &-0.399 C_{\phi l}^{(3)}[11]-0.399 C_{\phi l}^{(3)}[22]\bigg) \times 10^{-3} \ \textrm{GeV}
\end{split}
\end{align}
\begin{align}
\begin{split}
    \delta\Gamma&_\textrm{NLO}^{\overline{\mathrm{MS}}}(\hsm\rightarrow b\bar{b}) = -2.45 \times 10^{-3}\ \textrm{GeV} \\
    &+\left(\frac{1 \ \textrm{TeV}}{\Lambda}\right)^2 \bigg(-0.00103 C_{G}+0.00942 C_{\phi} +0.00294 C_{\phi B} -0.275 C_{\phi \square}+0.0701 C_{\phi D}\\
    &+1. C_{\phi G}+0.0113 C_{\phi W} +0.0388 C_{dB}[33]+0.0994 C_{dG}[33]+1.24 C_{d\phi }[33]+1.8 C_{dW}[33]\\
    &-0.00878 C_{ll}[1122]-0.15 C_{ll}[1221]+0.00118 C_{lq}^{(3)}[1133]+0.00118 C_{lq}^{(3)}[2233]\\
    &-0.00317 C_{\phi d}[33]+0.144 C_{\phi l}^{(3)}[11]+0.144 C_{\phi l}^{(3)}[22]+0.00273 C_{\phi q}^{(1)}[33]-0.023 C_{\phi q}^{(3)}[33]\\
    &-0.0363 C_{\phi ud}[33]-0.00218 C_{qd}^{(1)}[3333]-0.00231 C_{qd}^{(8)}[3333]+0.00184 C_{qq}^{(1)}[3333]\\
    &-0.00882 C_{qq}^{(3)}[3333]+0.00119 C_{qu}^{(1)}[3333]+0.00262 C_{qu}^{(8)}[3333]-0.00108 C_{quqd}^{(1)}[2233]\\
    &+2.76 C_{quqd}^{(1)}[3333]+0.389 C_{quqd}^{(8)}[3333]+0.00428 C_{uG}[33]-0.0119 C_{u\phi }[33]\\
    &+0.0258 C_{uW}[33]\bigg) \times 10^{-3} \ \textrm{GeV}
\end{split}
\end{align}
where we have truncated at $10^{-6}$ GeV.

For $\hsm \to c\bar{c}$, we find
\begin{align}
\begin{split}
    \Gamma^{\overline{\mathrm{MS}}}_\textrm{LO}&(\hsm\rightarrow c\bar{c}) = 2.28 \times 10^{-3} \ \textrm{GeV}\\
    &+\left(\frac{1 \ \textrm{TeV}}{\Lambda}\right)^2\bigg(2.28 +0.277 C_{\phi \square}-0.0691 C_{\phi D}+0.138 C_{ll}[1221]\\
    &-0.138 C_{\phi l}^{(3)}[11]-0.138 C_{\phi l}^{(3)}[22]-15.8 C_{u\phi }[22]\bigg) \times 10^{-3} \ \textrm{GeV}
\end{split}
\end{align}
\begin{align}
\begin{split}
    \delta\Gamma&_\textrm{NLO}^{\overline{\mathrm{MS}}}(\hsm\rightarrow c\bar{c}) = -1.01 \times 10^{-3}\ \textrm{GeV} \\
    &+\left(\frac{1 \ \textrm{TeV}}{\Lambda}\right)^2 \bigg(0.00328 C_{\phi} +0.00695 C_{\phi B} -0.114 C_{\phi \square}+0.0309 C_{\phi D}+0.45 C_{\phi G}\\
    &+0.00579 C_{\phi W} -0.00309 C_{ll}[1122]-0.0587 C_{ll}[1221]+0.0565 C_{\phi l}^{(3)}[11]\\
    &+0.0565   C_{\phi l}^{(3)}[22]+0.00296 C_{\phi q}^{(3)}[22]-0.00165 C_{qq}^{(3)}[2233]-1.42 C_{qu}^{(1)}[2332]\\
    &-1.98 C_{qu}^{(8)}[2332]-0.0186 C_{uB}[22]-0.0918 C_{uG}[22]+1.24 C_{u\phi }[22]\\
    &-0.00617 C_{u\phi }[33]+0.0544 C_{uW}[22]\bigg) \times 10^{-3} \ \textrm{GeV}
\end{split}
\end{align}
where we have truncated at $10^{-6}$.

Finally for $\hsm \to s\bar{s}$ we find
\begin{align}
\begin{split}
    \Gamma^{\overline{\mathrm{MS}}}_\textrm{LO}&(\hsm\rightarrow s\bar{s}) = 10 \times 10^{-6} \ \textrm{GeV}\\
    &+\left(\frac{1 \ \textrm{TeV}}{\Lambda}\right)^2\bigg(1.22 C_{\phi \square}-0.304 C_{\phi D}-1050. C_{d\phi }[22]+0.608 C_{ll}[1221]\\
    &-0.608 C_{\phi l}^{(3)}[11]-0.608 C_{\phi l}^{(3)}[22]\bigg) \times 10^{-6} \ \textrm{GeV}
\end{split}
\end{align}
\begin{align}
\begin{split}
    \delta\Gamma&_\textrm{NLO}^{\overline{\mathrm{MS}}}(\hsm\rightarrow s\bar{s}) = -8.49 \times 10^{-6}\ \textrm{GeV} \\
    &+\left(\frac{1 \ \textrm{TeV}}{\Lambda}\right)^2 \bigg(0.0145 C_{\phi} +0.0194 C_{\phi B} -0.995 C_{\phi \square}+0.254 C_{\phi D}+5.2 C_{\phi G}\\
    &+0.0229 C_{\phi W} +1.25 C_{dB}[22]-2.77 C_{dG}[22]+298. C_{d\phi }[22]+4.19 C_{dW}[22]\\
    &-0.00895 C_{ledq}[3322]-0.0137 C_{ll}[1122]-0.498 C_{ll}[1221]-0.00462 C_{\phi d}[22]\\
    &+0.489 C_{\phi l}^{(3)}[11]+0.489 C_{\phi l}^{(3)}[22]+0.00862 C_{\phi q}^{(1)}[22]+0.0176 C_{\phi q}^{(3)}[22]\\
    &-0.00384 C_{\phi q}^{(3)}[33]-0.0436 C_{\phi ud}[22]-0.00239 C_{qd}^{(1)}[2222]-0.0348 C_{qd}^{(1)}[2332]\\
    &-0.00335 C_{qd}^{(8)}[2222]-0.0445 C_{qd}^{(8)}[2332]-0.0062 C_{qq}^{(3)}[2233]-0.00244 C_{qq}^{(3)}[2332]\\
    &+0.00292 C_{qu}^{(1)}[3333]+0.00383 C_{qu}^{(8)}[3333]+0.134 C_{quqd}^{(1)}[2222]+45.1 C_{quqd}^{(1)}[2332]\\
    &+114. C_{quqd}^{(1)}[3322]+0.0278 C_{quqd}^{(8)}[2222]+19.1 C_{quqd}^{(8)}[2332]+1.17 C_{quqd}^{(8)}[3322]\\
    &+0.00163 C_{uG}[33]-0.0271 C_{u\phi }[33]\bigg) \times 10^{-6} \ \textrm{GeV}
\end{split}
\end{align}
where we have truncated at $10^{-9}$ GeV.

%% file: hvv.tex
\subsection{\texorpdfstring{$\hsm\rightarrow \gamma \gamma$}{H to gamma gamma}}\label{sec:apphvv}
With the inputs of Sec.~\ref{sec:inputs},
\begin{equation}
    \Gamma_\textrm{LO}(\hsm\rightarrow \gamma \gamma) = \left(\frac{1 \ \text{TeV}}{\Lambda}\right)^2 \left(-0.473{C_{\phi B}} -0.136 {C_{\phi W}}+0.253{C_{\phi WB}}\right)\times 10^{-3} \ \textrm{GeV}
\end{equation}
\begin{align}
\begin{split}
    \delta\Gamma_\textrm{NLO}&(\hsm\rightarrow \gamma\gamma) = 9.80 \times 10^{-6} \ \textrm{GeV} \\ &
    +\left(\frac{1 \ \textrm{TeV}}{\Lambda}\right)^2\bigg(4.78 C_{\phi B}+1.19 C_{\phi\square}-2.37 C_{\phi D}  -6.16 C_{\phi W}\\ &+3.08 C_{\phi WB}-0.333 C_{W}+1.78 C_{ll}[1221]-1.78 C_{\phi l}^{(3)}[11] -1.78 C_{\phi l}^{(3)}[22]\\ &+13.5 C_{uB}[33]+0.338 C_{u\phi}[33]+7.22 C_{uW}[33] \bigg) \times 10^{-6} \ \textrm{GeV}
\end{split}
\end{align}

\subsection{\texorpdfstring{$\hsm\rightarrow \gamma Z$}{H to gamma Z}}

With the inputs of Sec.~\ref{sec:inputs} and taking all fermions massless except for $m_t$,
\begin{equation}
    \Gamma_\textrm{LO}(\hsm\rightarrow  \gamma Z) = \left(\frac{1 \ \textrm{TeV}}{\Lambda}\right)^2\left(0.0939 C_{\phi B} -0.0939 C_{\phi W}+0.0625 C_{\phi WB}\right) \times 10^{-3} \ \textrm{GeV}\, .
\end{equation}
\begin{align}
\begin{split}
    \delta\Gamma_\textrm{NLO}&(\hsm\rightarrow \gamma Z) = 6.52 \times 10^{-6} \ \textrm{GeV} \\
    &+ \left(\frac{1 \ \textrm{TeV}}{\Lambda}\right)^2\bigg(2.08 C_{\phi B}+0.791 C_{\phi \square}-0.785 C_{\phi D}-2.52 C_{\phi W} \\
    & -2.31 C_{\phi WB}-0.493 C_W+1.19 C_{ll}[1221]-1.19 C_{\phi l}^{(3)}[11]-1.19 C_{\phi l}^{(3)}[22] \\
    & +0.115 C_{\phi q}^{(1)}[33]-0.115 C_{\phi q}^{(3)}[33]+0.115 C_{\phi u}[33]-0.457 C_{uB}[33] \\ 
    &+0.0471 C_{u\phi}[33]+3.00 C_{uW}[33] \bigg) \times 10^{-6} \ \textrm{GeV}\, .
\end{split}
\end{align}

\subsection{\texorpdfstring{$\hsm\rightarrow gg$}{H to gg}}
With the inputs of Sec.~\ref{sec:inputs} and the assumptions discussed in Sec. \ref{sec:htovv},
\begin{equation}
    \Gamma_\textrm{LO}(\hsm\rightarrow  gg) = 7.60 C_{\phi G} \left(\frac{1 \ \textrm{TeV}}{\Lambda}\right)^2\times 10^{-3} \ \textrm{GeV}
\end{equation}
\begin{align}
\begin{split}
    \delta\Gamma_\textrm{NLO}&(\hsm\rightarrow gg) = 0.193 \times 10^{-3} \ \textrm{GeV} \\ 
    &+\left(\frac{1 \ \textrm{TeV}}{\Lambda}\right)^2\bigg(0.0234 C_{\phi \square }-0.00585 C_{\phi D}+0.0130 C_{dG}[33] \\
    &+0.0469 C_{d\phi}[33]+0.0117 C_{ll}[1221]-0.0117 C_{\phi l}^{(3)}[11]-0.0117 C_{\phi l}^{(3)}[22] \\
    &-0.189 C_{uG}[33]-0.0249 C_{u\phi}[33] \bigg) \times 10^{-3} \ \textrm{GeV}\, .
\end{split}
\end{align}

%% file: terasens.tex
We report here the numerical values used for the observables entering the results presented in this work.
The current values of the electroweak precision observables used in Sec.~\ref{sec:hl-lhc} are summarized in Tab.~\ref{tab:expnums}. The experimental values are taken from Tab. 10.3 of the Particle Data Group~\cite{PhysRevD.110.030001}, while the theoretical predictions include the full set of two-loop contributions
and higher order corrections when known. The theory predictions are computed
using the formulae in the indicated references and the input parameters of Sec.~\ref{sec:inputs}, and the theory errors
include the parametric uncertainties on the top and Higgs masses~\cite{Dubovyk:2019szj}, along with the estimated
theory uncertainties described in the respective papers.

\begin{table}
\begin{center}
\begin{tabular}{|l|c|c|}
\hline
Measurement& Current Experiment& Current  theory
\\
\hline\hline
$\Gamma_Z$(GeV) & $2.4955\pm 0.0023$ & $2.4943\pm 0.0006$   \cite{Freitas:2014hra,Dubovyk:2018rlg,Dubovyk:2019szj} 
\\
\hline
$R_e$ & $20.804\pm 0.05$ &  $20.732\pm 0.009 $ \cite{Freitas:2014hra,Dubovyk:2018rlg,Dubovyk:2019szj} \\
\hline
$R_\mu$ & $20.784\pm 0.034$ & $20.732\pm 0.009 $ \cite{Freitas:2014hra,Dubovyk:2018rlg,Dubovyk:2019szj}  \\
\hline 
$R_\tau$ & $20.764\pm 0.045$ & $ 20.779\pm 0.009$ \cite{Freitas:2014hra,Dubovyk:2018rlg,Dubovyk:2019szj}\\
\hline 
$R_b$& $0.21629\pm 0.00066$ & $ 0.2159\pm 0.0001$~\cite{Freitas:2014hra,Dubovyk:2018rlg,Dubovyk:2019szj}   \\
\hline
$R_c$ & $0.1721 \pm 0.0030$ & $ 0.1722\pm 0.00005 $~\cite{Freitas:2014hra,Dubovyk:2018rlg,Dubovyk:2019szj} \\
\hline
$\sigma_h$ & $41.4802\pm 0.0325 $ & $41.492\pm 0.008 $~\cite{Freitas:2014hra,Dubovyk:2018rlg,Dubovyk:2019szj}    
\\
\hline 
$A_e ({\text{from}}~A_{LR}~{\text{had}}$ & $0.15138\pm 0.00216$ & $ 0.1469\pm 0.0004$ \cite{Dubovyk:2019szj,Awramik:2006uz}\\
\hline
$A_e ({\text{from}}~A_{LR}~{\text{lep}})$ & $0.1544\pm 0.0060$& $ 0.1469\pm 0.0004$ \cite{Dubovyk:2019szj,Awramik:2006uz}\\
\hline
$A_e ({\text{from~Bhabba~pol}})$ & $0.1498\pm 0.0049$ & $ 0.1469\pm 0.0004$ \cite{Dubovyk:2019szj,Awramik:2006uz}\\
\hline
$A_\mu$ & $0.142\pm 0.015$ & $ 0.1469\pm 0.0004$ \cite{Dubovyk:2019szj,Awramik:2006uz}\\
\hline
$A_\tau ({\text{from~SLD}})$ & $0.136\pm 0.015$ & $ 0.1469\pm 0.0004$ \cite{Dubovyk:2019szj,Awramik:2006uz}\\
\hline
$A_\tau (\tau~ {\text{pol}})$ & $0.1439\pm 0.0043$ & $ 0.1469\pm 0.0004$ \cite{Dubovyk:2019szj,Awramik:2006uz}\\
\hline
$A_c$ & $0.670\pm 0.027$ & $ 0.66773\pm 0.0002$~\cite{Dubovyk:2019szj,Awramik:2006uz} \\
\hline 
$A_b$ & $0.923\pm 0.020$ & $0.92694\pm 0.00006$~\cite{Dubovyk:2019szj,Awramik:2006uz,Awramik:2008gi} 
 \\\hline  
$A_s$ & $0.895\pm 0.091$ & $  0.93563\pm 0.00004$~\cite{Dubovyk:2019szj,Awramik:2006uz}
\\
 \hline  
$A_{e,FB}$ & $0.0145\pm 0.0025$ & $ 0.0162\pm 0.0001$ \cite{Dubovyk:2019szj,Awramik:2006uz}\\
 \hline
$A_{\mu,FB}$ & $0.0169\pm 0.0013$ & $ 0.0162\pm 0.0001$ \cite{Dubovyk:2019szj,Awramik:2006uz} \\
 \hline
 $A_{\tau,FB}$ & $0.0188\pm 0.0017$ & $ 0.0162\pm 0.0001$ \cite{Dubovyk:2019szj,Awramik:2006uz} \\
 \hline
$A_{b,FB}$ & $0.0996\pm 0.0016$ & $ 0.1021\pm 0.0003$ \cite{Dubovyk:2019szj,Awramik:2006uz,Awramik:2008gi} \\
\hline
$A_{c,FB}$ & $0.0707\pm 0.0035$ & $ 0.0736\pm 0.0003$  \cite{Dubovyk:2019szj,Awramik:2006uz}\\
\hline
$A_{s,FB}$ & $0.0976\pm 0.0114$ & $0.10308 \pm 0.0003$ \cite{Dubovyk:2019szj,Awramik:2006uz} \\
\hline
$\Gamma_W$(GeV)  & $2.085\pm 0.042$& $ 2.0903\pm  0.002$~\cite{deBlas:2944678}\\
\hline
$\alpha(M_Z^2)^{-1} $ & $127.93$ & $127.93\pm 0.014$~\cite{ParticleDataGroup:2024cfk} \\ \hline
\end{tabular}
\caption{EWPO used in numerical results of this work. Experimental results are taken from Tab. 10.3 of~\cite{PhysRevD.110.030001}.
The theory results include the full set of two-loop contributions
for the $Z$ pole observables, along with higher order corrections when known.     The theory predictions are
computed using the formulae in the indicated references and our input parameters as detailed in the text. \label{tab:expnums}}
\end{center}
\end{table}

For the FCC-ee projections, we provide the relative percentage uncertainty on Higgs production via $e^+ e^- \to Zh$, followed by each decay channel, in Tab.~\ref{tab:fcc}. We consider two energy runs, at $\sqrt{s}=240$ GeV with  statistics of $10.8$ ab$^{-1}$ and at $\sqrt{s}=240$ GeV with $3.12$ ab$^{-1}$~\cite{Altmann:2025feg}.   For the $W W^*$ decays, we include the fully hadronic $W$ decays, with the $Z$ decaying to $e^+e^-,\mu^+\mu^-$ or $\nu {\overline{\nu}}$.

\begin{table}[t!]
\begin{center}
\begin{tabular}{|c|c|c|}
\hline
 & $\sqrt{s}=240~\textrm{GeV}$ & $\sqrt{s}=365~\textrm{GeV}$\\
 \hline\hline
 $b\overline{b}$ & 0.21 & 0.38\\
 \hline 
 $c\overline{c}$ & 1.6 & 2.9\\
 \hline 
 $s\overline{s}$ & 120 & 350\\
 \hline 
 $gg$ & 0.8 & 2.1\\
 \hline 
 $\tau^+\tau^-$ & 0.58 & 1.2\\
 \hline 
 $\mu^+\mu^-$ & 11 & 25\\
 \hline 
 $W W^*$ & 0.8 & 1.8\\
 \hline 
 $ZZ^*$ & 2.5 & 8.3\\
 \hline 
 $\gamma\gamma$ & 3.6 & 13\\
 \hline
 $Z\gamma$ & 11.8 & 22\\
 \hline
\end{tabular}
\caption{Uncertainty in $\%$ on Higgstrahlung measurements  including Higgs decays  at FCC-ee with $10.8$~ab$^{-1}$ ($3.12$~ab$^{-1}$) at $\sqrt{s}=240$ GeV ($\sqrt{s}=365$ GeV)~\cite{Selvaggi:2025kmd}.}
\label{tab:fcc}
\end{center}
\end{table}

\begin{table}
\begin{center}
\begin{tabular}{|l|c|c|c|c|c|c|}
\hline
FCC-ee&&&&&&\\
Uncertainties& Stat& Syst & PO(C)&PO(A)&Theory (C)&Theory (A)
\\
\hline\hline
$\Delta\Gamma_Z$(KeV) & $4$ &  $12$   &$35$&$-$&$80$& $16$
\\
\hline
$\delta R_e$ & $3.4 \times 10^{-6}$~\cite{Selvaggi:2025kmd} & $2.3\times 10^{-6}$~\cite{Selvaggi:2025kmd} & $4 \times 10^{-4}$ &$-$&$1.2\times 10^{-3}$&$2\times 10^{-4}$\\
\hline
$\delta R_\mu$ & $2.4\times 10^{-6}$ & $2.3\times 10^{-6}$   & $4\times 10^{-4}$&$-$&
$1.2\times 10^{-3}$&$2\times 10^{-4}$\\
\hline 
$\delta R_\tau$ & $2.7\times 10^{-6}$~\cite{Selvaggi:2025kmd} & $ 2.3\times 10^{-6}$~\cite{Selvaggi:2025kmd} & $4\times 10^{-4}$&$-$&$1.2\times 10^{-3}$  &$2\times 10^{-4}$\\
\hline 
$\delta R_b$& $ 1.2 \times 10^{-6}$ & $ 1.6\times 10^{-6}$  &$4.4\times 10^{-5}$&$9 \times 10^{-6}$&$2\times 10^{-5}$&$3.5\times 10^{-6}$\\
\hline
$\delta R_c$ &$1.4\times 10^{-6}$~\cite{Selvaggi:2025kmd} &  $2.2\times 10^{-6}$~\cite{Selvaggi:2025kmd} & $1.7\times 10^{-4}$&$3.4 \times 10^{-5}$& $1\times 10^{-5}$&$2\times 10^{-6}$\\
\hline
$\Delta \sigma_h (pb)$ & 0.03\cite{Selvaggi:2025kmd} &  0.8\cite{Selvaggi:2025kmd} 
& $1.7$&$-$&$1.6$&$0.3$\\
\hline 
$\Delta A_e$  & $14 \times 10^{-6}$~\cite{Selvaggi:2025kmd} & $11 \times 10^{-6}$~\cite{Selvaggi:2025kmd}&$19.5\times 10^{-5}$ [*]&$-$& $5.3 \times 10^{-5}$ [*] & $4.5 \times 10^{-6}$ [*]\\
\hline
$\Delta A_\mu$ & \multicolumn{2}{c|}{$32\times 10^{-6}$~\cite{Selvaggi:2025kmd}} & $19.5\times 10^{-5}$ [*]&$-$ & $5.3 \times 10^{-5}$ [*]& $4.5 \times 10^{-6}$ [*]\\
\hline
$\Delta A_\tau$  & \multicolumn{2}{c|}{$34\times 10^{-6}$~\cite{Selvaggi:2025kmd}} & $19.5\times 10^{-5}$ [*]&$-$& $5.3 \times 10^{-5}$ [*]& $4.5 \times 10^{-6}$ [*]\\
\hline
$\Delta A_c$ &\multicolumn{2}{c|}{$60\times 10^{-6}$~\cite{Selvaggi:2025kmd}} &$91\times 10^{-5}$ [*]& $-$ & $2.3 \times 10^{-5}$ [*] & $2 \times 10^{-6}$ [*] \\
\hline 
$\Delta A_b$ & \multicolumn{2}{c|}{$98\times 10^{-6}$~\cite{Selvaggi:2025kmd}}
&$126\times 10^{-5}$[*]&$-$  & $4.3 \times 10^{-6}$ [*] & $3.7\times 10^{-7}$ [*] \\\hline  
 \hline  
$\Delta A_{e,FB}$ & $3.3\times 10^{-6}$~\cite{Selvaggi:2025kmd} & $2.4\times 10^{-6}$~\cite{Selvaggi:2025kmd}  &$4.3 \times 10^{-5}$&$-$& $1.2 \times 10^{-5}$ [*] & $1 \times 10^{-6}$ [*]\\
 \hline
 $\Delta A_{\mu,FB}$ & $2.3\times 10^{-6}$~\cite{Selvaggi:2025kmd} & $2.4\times 10^{-6}$~\cite{Selvaggi:2025kmd}  &$4.3 \times 10^{-5}$&$-$&$1.2 \times 10^{-5}$ [*] & $1 \times 10^{-6}$ [*]\\
 \hline
 $\Delta A_{\tau,FB}$ & $2.8\times 10^{-6}$~\cite{Selvaggi:2025kmd} & $2.4\times 10^{-6}$~\cite{Selvaggi:2025kmd}  &$4.3 \times 10^{-5}$&$-$& $1.2 \times 10^{-5}$ [*] & $1 \times 10^{-6}$ [*]\\
 \hline
$\Delta A_{b,FB}$ & $4\times 10^{-6}$~\cite{Selvaggi:2025kmd}&  $4\times 10^{-6}$~\cite{Selvaggi:2025kmd}  &$3.2 \times 10^{-5}$& $2.8\times 10^{-6}$& $3.8 \times 10^{-5}$ [*] & $3.2 \times 10^{-6}$ [*] \\
\hline
$\Delta A_{c,FB}$ & $5\times 10^{-6}$~\cite{Selvaggi:2025kmd}&  $5\times 10^{-6}$~\cite{Selvaggi:2025kmd}   &$2.3 \times 10^{-5}$&$2.1\times 10^{-6}$& $2.9 \times 10^{-5}$ [*] & $2.5 \times 10^{-6}$ [*]\\
\hline
$\Delta\Gamma_W$(KeV)  & $270$& $ 200$ &&&$1000$&$100$\\
\hline
$\Delta \alpha(M_Z)^{-1}$ ~& $8\times 10^{-4}$& $3.8\times 10^{-3}$&&&$5\times 10^{-5}$&$2\times 10^{-5}$\\ \hline
\end{tabular}
\caption{Projected uncertainties on EWPO at FCC-ee  from Tera-Z with 205 $ab^{-1}$ and $\Gamma_W$ measured at the $WW$ threshold with 19.2 $ab^{-1}$, from Ref.~\cite{deBlas:2944678} unless otherwise stated. Theory uncertainites are divided into those associated with the definition of pseudo-observables at the $Z$ peak (``PO'') and those coming from calculations (``Theory''). Both conservative (``C'') and aggressive (``A'') scenarios are considered. The dash means the uncertainty is assumed to be negligible. Numbers labeled by [*] are found by linear propagation of errors from Ref.~\cite{deBlas:2944678}. \label{tab:fcceenums}}
\end{center}
\end{table}

Tab.~\ref{tab:fcceenums} reports the projected uncertainties on the EWPO at FCC-ee. All numbers are taken from~ Ref.~\cite{deBlas:2944678}, unless otherwise stated. For each observable, we detail the statistical, systematic and theoretical errors, and add them in quadrature for the results of this work. We further divide the theoretical uncertainties between those associated with the extraction of the pseudo-observables from the data at the $Z$ pole, which requires theory inputs for instance on the simulation of radiation and modeling of hadronization and other non-perturbative effects (labeled ``PO''), and those that directly come from perturbative calculations (``Theory''). We consider two possible future  scenarios: a conservative one (``C''), where theory improvements are reached only by extending existing computational methods, and an aggressive one (``A''), where more fundamental advances in theoretical techniques is foreseen. 

The dash indicates a negligible uncertainty as compared to other error sources. Entries where the information separation between ``PO'' and ``Theory'' uncertainties does not apply have been left blank, while the symbol [*] denotes that the number has been obtained by linear propagation of errors, following a similar approach as Ref.~\cite{Maura:2024zxz}.
We denote relative uncertainties with $\delta$ and absolute ones with $\Delta$.